\documentclass[draftclsnofoot, onecolumn]{IEEEtran}
\usepackage{graphicx}
\usepackage{amstext}
\usepackage{amsmath}
\usepackage{amssymb}

\begin{document}

\newtheorem{Thm}{Theorem}[section]
\newtheorem{Cor}[Thm]{Corollary}
\newtheorem{Lem}[Thm]{Lemma}
\newtheorem{Propn}[Thm]{Proposition}
\newtheorem{Def}[Thm]{Definition}
\newtheorem{rem}[Thm]{Remark}
\newtheorem{Asmp}[Thm]{Assumption}

\title{ 1-Bit Compressive Sensing: Reformulation and RRSP-Based Sign Recovery Theory}
\author{Yun-Bin Zhao\thanks{Corresponding author. School of
Mathematics, University of Birmingham,  Edgbaston,  Birmingham, West Midlands  B15
2TT,  United Kingdom (e-mail: {\tt y.zhao.2@bham.ac.uk}).  The research of
this author was supported by the Engineering and Physical Sciences
Research Council (EPSRC) under  grant \#EP/K00946X/1. } ~ and
Chunlei Xu\thanks{School of Mathematics, University of Birmingham,
  Birmingham, West Midlands  B15 2TT,  United Kingdom (e-mail: {\tt
xuc@for.mat.bham.ac.uk}). } }

\maketitle

\begin{abstract} Recently, the 1-bit compressive sensing (1-bit CS) has been
 studied in the field of sparse signal recovery. Since the amplitude
information of sparse signals in 1-bit CS is not available,   it is
often the support or the sign   of a signal that can be exactly
recovered with a decoding method. In this paper, we  first show that
a necessary assumption (that has been overlooked in the literature)
   should be made for some existing theories and discussions for 1-bit
  CS. Without such an assumption, the found solution by some existing decoding algorithms might be inconsistent with 1-bit measurements.   This motivates us
    to pursue a new direction to develop  uniform and nonuniform recovery theories for
     1-bit CS  with a new decoding method which always generates a solution consistent with 1-bit measurements.
   We focus on an extreme case of 1-bit CS, in which the measurements capture only the sign of the
    product of a sensing matrix and a signal.
     We show that the 1-bit CS model can be reformulated equivalently as
    an $\ell_0$-minimization problem with linear constraints.  This reformulation naturally
    leads to a new linear-program-based decoding method, referred to
      as the
   1-bit basis pursuit, which is remarkably different from existing formulations.
   It turns out that the uniqueness condition for the solution
   of the 1-bit basis pursuit  yields the so-called  restricted range
   space property (RRSP) of the transposed  sensing matrix.  This concept provides a basis to develop
sign recovery conditions for sparse signals
   through
    1-bit  measurements. We prove that if the sign of a sparse signal can be exactly recovered
    from 1-bit measurements with 1-bit
     basis pursuit, then the sensing matrix must admit
     a certain RRSP, and that if the sensing matrix admits a slightly enhanced
     RRSP, then the sign of a $k$-sparse signal can be exactly recovered
 with  1-bit basis pursuit.
 \end{abstract}

{\IEEEkeywords  1-bit compressive sensing,
  restricted range space property, 1-bit basis
pursuit, linear program, $\ell_0$-minimization, sparse signal
recovery.}

\section{Introduction}

Compressive sensing (CS) has attracted plenty of recent attention in
the field of signal and image processing.    One of the key
mathematical issues addressed in CS  is how
  a sparse signal can be reconstructed  by a   decoding algorithm. An extreme case of CS  can be cast as the problem of seeking the sparsest solution of
an underdetermined linear system, i.e.,
$$\min\{\|x\|_0:~ \Phi x=b\},$$ where $\|x\|_0$ counts the number of nonzero components of $x$,  $\Phi\in R^{m\times n}$
($m<n$) is called a sensing matrix,  and $b\in R^m$ is the vector of
nonadaptive measurements.  It is known that the reconstruction of
a sparse signal   from a reduced number of acquired measurements is
possible when the sensing matrix $\Phi$ admits certain properties (see,
e.g., \cite{DE2003,T2004,ET2005, ERT2006, ERTR2006, D2006, CDD2009,
 YZ2008,YBZ2013, FR2013}). Note that measurements must be quantized.
 Fine quantization provides more information on a signal, making the signal more likely to be  exactly recovered.
  However,  fine quantization  imposes a huge burden on measurement systems,
  leading to slower sampling rates and increased costs for hardware systems
  (see, e.g. \cite{W99, LRRB05, SG09, B2010}). Also, fine quantization introduces error to   measurements.
  This motivates one to consider sparse signal recovery through lower bits of measurements. An extreme
  quantization is only one bit per measurement.
 As demonstrated in
\cite{BB2008,B2009} and \cite{B2010},  it is possible, in some
situations, to reconstruct a sparse signal within certain factors
from 1-bit measurements, e.g., the sign of measurements. This
motivates the recent development of CS with 1-bit measurements,
called 1-bit compressive sensing (see, e.g., \cite{BB2008,B2009,
GNR2010, L2011,LWYB2011,LB2012, PV20138}). An ideal model for 1-bit
CS  is the $\ell_0$-minimization with sign constraints
\begin{eqnarray}\label{1bitCS}
\min\{\|x\|_0:~\textrm{sign}(\Phi x)=y\},
\end{eqnarray}
where $\Phi\in R^{m\times n}$ is a sensing matrix and $y\in R^m$ is
the vector of 1-bit measurements. Throughout the paper, we assume
that $m<n.$ The sign function in (\ref{1bitCS})  is applied
element-wise. Due to the NP-hardness of (\ref{1bitCS}), some
relaxations of (\ref{1bitCS}) have been investigated in the
literature. A common relaxation is replacing $\|x\|_0$ with
$\|x\|_1$ and replacing the constraint of (\ref{1bitCS}) with the
linear system
\begin{equation}\label{YYY} Y \Phi x \geq 0, \end{equation}   where $ Y=\textrm{diag} (y). $ In addition, an
extra constraint, such as $\|x\|_2 =1$ and $\|\Phi x\|_1=m,$ is
 introduced into this relaxation model in order to exclude some
trivial solutions.

Only the acquired 1-bit information   is insufficient to exactly reconstruct a sparse signal.
 For instance, if  $\textrm{sign} (\Phi x^*) = y$ where $y \in \{1, -1\}^m,$  then any small perturbation $x^*+ u$
 also satisfies this equation, making the exact recovery of $x^*$ almost impossible by
 whichever decoding algorithms.  While the sign
information of measurements might not be enough to exactly
reconstruct a signal, it might be adequate to recover the support or the sign  of the signal. Thus 1-bit CS still has found applications in
signal recovery
 \cite{BB2008,B2009, GNR2010, B2010, L2011},   imaging processing \cite{BAU2010,BU2013}, and   matrix completion
\cite{DPBW2014}.

The 1-bit CS was first proposed and investigated by Boufounos and
Baraniuk \cite{BB2008}.  Since 2008,  numerous algorithms  have been
developed in this direction, including greedy algorithms (see, e.g.,
\cite{B2009,GNR2010,KBAU2012,YYO2012, JLBB2013,GNJN2013, BBR2013})
and convex and nonconvex programming algorithms (see, e.g.,
\cite{BB2008,LWYB2011,MPD2012, PV20131,PV20138,  SS2013, ALPV2014}).
To find a polynomial-time solver for the 1-bit CS problems, a linear programming
model  based on (\ref{YYY})  has been formulated, and certain stability
results for reconstruction have been shown in \cite{PV20138} as
well.

 In classic
CS setting, it is well known that when a sensing matrix
admits some properties such as  mutual coherence \cite{DE2003,
BED2009}, null space property (NSP) \cite{CDD2009,YZ2008}, restricted
isometry property (RIP) \cite{ET2005} or  range space property
(RSP) of $\Phi^T$ \cite{YBZ2013}, the  signals with
low sparsity levels
  can be exactly recovered  by the basis pursuit
  and other algorithms.   This motivates one  to investigate whether
  similar recovery theories can also be established for 1-bit CS problems.
In  \cite{JLBB2013}, the binary iterative hard thresholding (BIHT)
algorithm  for 1-bit CS problems is discussed and the so-called
binary $\varepsilon$-stable embedding (B$\epsilon$SE) condition is
introduced.   The B$\epsilon$SE   can be seen as an
extension of the
 RIP. However, at the current stage, the theoretical analysis for the guaranteed
performance of 1-bit CS algorithms is far from complete,
in contrast to the classic CS.
 Recovery conditions in terms of the property of $\Phi$ and/or $y$ are still under development.

 The fundamental assumption on 1-bit CS is that any solution $x$ generated by an algorithm should
 be consistent with the acquired 1-bit measurements   in the sense that
 \begin{equation} \label{consistency} \textrm{sign}(\Phi x) = y = \textrm{sign}(\Phi x^*),
 \end{equation} where $x^*$ is the targeted signal.
 Clearly, it is very difficult to directly solve a problem with such a
 constraint
  if it does not have a tractable reformation.    From a computational point of view, an ideal
  relaxation or reformulation of the sign constraint is a linear system.    The current algorithms and theories for 1-bit CS
  (e.g., \cite{BB2008, B2010,PV20138, SS2013}) have been  developed largely based on the system (\ref{YYY}), which
  is a linear relaxation of (\ref{consistency}).    In Section II of this paper,   we show that
  the existing relaxation  based on (\ref{YYY})   is not equivalent to the original 1-bit CS model.
  In fact, a vector satisfying (\ref{YYY}) together with a trivial-solution excluder, such as $\|x\|_2=1$ or $\|\Phi x\|_1=m,$
   may
  not be consistent with the acquired 1-bit measurements $y. $
  Some necessary conditions  must be imposed on the matrix in order to ensure that the solution of a decoding algorithm
  based on (\ref{YYY}) is consistent with $y.$ These necessary conditions  have been overlooked in the literature (see the discussion in Section II for details).

Many existing discussions for 1-bit CS do not distinguish between
zero and positive measurements. Both are mapped to 1 (or $-1$) by a
nonstandard sign function. In Section II, we point out that it is
beneficial to allow $y$ admitting zero components and to treat zero
and nonzero measurements separately from both practical
  and  mathematical points of view. Failing to distinguish zero and
  nonzero magnitude
 of measurements might yield  ambiguity of measurements when
  sensing vectors are nearly orthogonal to the signal. Such ambiguity might prevent
  from acquiring a correct sign of measurements due to signal noises or errors in computation.

This motivates us  to pursue a new direction to establish a recovery theory for
 1-bit CS. Our study is remarkably different from existing ones in several
 aspects.

 (a) The acquired sign measurements $y$    is allowed to admit zero
components. When $y$ does not contain zero components,
 our model immediately reduces to the existing 1-bit CS model.

(b) We introduce a truly equivalent reformulation of the 1-bit CS
model (\ref{1bitCS}). The  model
(\ref{1bitCS}) is reformulated equivalently as an
$\ell_0$-minimization problem with linear constraints. Replacing
$\|x\|_0$ with $\|x\|_1$  leads naturally to a new linear-program-based decoding
method, referred to as the 1-bit basis pursuit. Different from existing formulations,   the new
 reformulation ensures that the solution of the 1-bit basis pursuit
 is always consistent with the acquired 1-bit measurements
$y.$

(c) The   sign recovery theory developed in the paper is from the
  perspective of the restricted range space properties (RRSP)
of transposed sensing matrices.  In classic CS, it has been shown in
\cite{YBZ2013} that  any $k$-sparse signal can be exactly
recovered with  basis pursuit if and only if the transposed sensing
matrix admits the so-called range space property (RSP) of order $k.
$  This property   is equivalent to the well known NSP of order $k$ in the
sense that both are the
  necessary and sufficient ccondition for the  uniform recovery of
$k$-sparse signals.  The new reformulation of the 1-bit CS model
proposed in this paper makes it possible to develop an analogous
recovery guarantee for the sign of sparse signals with 1-bit basis
pursuit. This development naturally yields the concept of
the restricted range space property (RRSP) which   gives
rise to some necessary   and sufficient    conditions for the nonuniform
and uniform recovery of  the sign of sparse signals from 1-bit
measurements.

The main results of the paper can be summarized as follows:

\begin{itemize}
\item \emph{(Theorem \ref{Main36}, nonuniform)}  If the 1-bit basis pursuit can exactly recover the sign of
$k$-sparse signals consistent with 1-bit measurements $y , $ then
$\Phi$ must admit the N-RRSP of order $k $ with respect to $y $ (see
Definition \ref{NRRSP-y}).

\item \emph{(Theorem \ref{Main38}, nonuniform)} If $\Phi$ admits the  S-RRSP of order $k $
with respect to $y$  (see Definition \ref{SRRSP-y}), then from  1-bit measurements, the 1-bit
basis pursuit can exactly recover the sign of $k$-sparse signals
 which
are the sparsest vectors consistent with $y.$

\item \emph{(Theorem \ref{Uniform-1}, uniform)}  If the 1-bit basis pursuit can exactly recover the sign of
all $k$-sparse signals from 1-bit measurements, then $ \Phi $ must
admit the so-called N-RRSP of order $k $ (see Definition
\ref{N-RRSP}).

\item \emph{(Theorem  \ref{Main2}, uniform)} If the matrix admits the S-RRSP of order $k$  (see Definition
\ref{S-RRSP}), then from  1-bit measurements, the 1-bit basis
pursuit can exactly recover the sign of all $k$-sparse signals which
are the sparsest vectors consistent with 1-bit measurements.
\end{itemize}
 The  above-mentioned definitions and theorems
 are given in Sections III and VI.  Central to the proof of these results is
  Theorem 3.2 which provides a full characterization for
the uniqueness of solutions to the 1-bit basis pursuit, and thus
yields a fundamental basis to develop  recovery conditions.

 This paper is organized as follows. We provide motivations for a new
 reformulation of the 1-bit CS model in Section II.
Based on the reformulation, nonuniform sign recovery conditions
  with 1-bit basis pursuit are developed in Section
III, and uniform sign recovery conditions are developed in Section
IV. The proof of Theorem 3.2   is given in Section V.

 We use the following notation in the paper. Let $R_+^n$  be
the set of nonnegative  vectors in
$R^n. $ The vector $x\in R^n_+$ is also written as $x\geq
0.$  Given a set $S$, $|S|$ denotes the cardinality of
$S$. For   $x \in R^n $ and $S\subseteq \{1, \dots, n\},$ let
  $x_{S}\in
R^{|S|}$  denote the subvector of $x$ obtained by deleting those
 components $x_i$ with $i\notin S,$  and  let $\textrm{supp}(x)=\{i:x_i\neq 0\}$ denote the support of
$x.$    The $\ell_0$-norm $\|x\|_0 $ counts the number of nonzero
components of $x$, and the $\ell_1$-norm of   $x$ is defined as
$\|x\|_1=\sum_{i=1}^{n}|x_i|$. For a matrix $\Phi   \in R^{m\times
n},$  we use $\Phi^T$ to denote the transpose of $\Phi,$ ${\cal N}
(\Phi) =\{x: \Phi x=0\}$  the null space of $\Phi,$ ${\cal R}(
\Phi^T) = \{ \Phi^T u:  u\in R^m\}$  the range space of $\Phi^T,$
$\Phi_{J,n}$  the submatrix of $\Phi $ formed by deleting the rows
of $\Phi$ which are not indexed by $J,$ and $\Phi_{m,J}$ the
submatrix of $\Phi $ formed by deleting the columns of $\Phi$ which
are not indexed by $J$. $e$ with a suitable dimension is the vector
of ones, i.e., $ e= (1,\dots, 1)^T. $

\section{Reformulation of 1-bit compressive sensing}

 In this section, we  point out that
for a given matrix, existing 1-bit CS algorithms based on
the  relaxation (\ref{YYY}) cannot guarantee the found solution
being
  consistent with the acquired 1-bit measurements $y,$ unless the  matrix satisfies
  some
condition.  This motivates one to propose a new reformulation of the 1-bit CS problem so that the resulting algorithm can automatically ensure its solution being consistent with 1-bit measurements.

\subsection{Consistency conditions for existing 1-bit CS methods}

 The standard sign function  is   defined as  $\textrm{sign} (t) =1 $ if $ t
>0,$ $\textrm{sign} (t)=-1$ if $ t <0, $ and $\textrm{sign} (t) =0$  otherwise.
  In the 1-bit CS literature,  many researchers do not distinguish between zero and positive
values of measurements and thus define $\textrm{sign} (t) =1$ for
$t\geq 0$ and $\textrm{sign} (t)=-1 $ otherwise. The function
$\textrm{sign}(\cdot) $ defined this way is referred to as a
nonstandard sign function in this paper. We now point out that no
matter a standard or nonstandard sign function is used,
   \emph{the equation $  y = \textrm{sign} (\Phi x) $   is generally not equivalent to the system
   (\ref{YYY}) even if a trivial-solution excluder such as $\|x\|_2=1$ or $\|\Phi x\|_1 =m $ is used, unless
   certain necessary assumptions are made on $\Phi.$}
  First, since  $ y = \textrm{sign} (\Phi x)  $ implies $ Y \Phi x \geq 0 $ (this fact was observed in \cite {BB2008}),
  the following statement is obvious:

\vskip 0.05in

\begin{Lem} \label{L21}  If $\Phi \in R^{m\times n}$ and $y\in \{1, -1\}^m$ or $y\in \{1, 0, -1\}^m,$
then  $\{x:   \textrm{sign}(\Phi x)= y \} \subseteq \{x: Y\Phi x\geq
0\}.$
\end{Lem}

 \vskip 0.05in

Without a further assumption on $\Phi,$ however,   the system
(\ref{YYY}) does not imply $  \textrm{sign} ( \Phi x) =y $ even if
some trivial solutions of   (\ref{YYY})  are excluded by adding a
widely used trivial-solution excluder, such as $\|x\|_2=1$ or
$\|\Phi x\|_1 =m,$ to the system.  In fact, for any given $y$ with $
J_-= \{i: y_i =-1\} \not=\emptyset,$ we see that all vectors $0\not=
\widetilde{x} \in {\cal N}(\Phi ) $ (or more generally,
$\widetilde{x} \not =0$ satisfying $  \Phi_{J_-, n} \widetilde{x} =0
$ and $ \Phi_{J_+, n} \widetilde{x} \geq 0 $) satisfy $Y \Phi
\widetilde{x} \geq 0, $ but for these vectors, $ \textrm{sign} (
\Phi \widetilde{x}) \not = y$ no matter $\textrm{sign} (\cdot)$  is
standard or nonstandard. The trivial-solution excluder   $\|x\|_2 =1
$ (e.g., \cite{BB2008}) cannot exclude  vectors satisfying $0\not=
\widetilde{x} \in {\cal N}(\Phi ) $    from the set $\{x: Y \Phi
x\geq 0\}.$ The excluder $\|\Phi x\|_1=m$ (e.g., \cite
{PV20138,SS2013}) cannot exclude $\widetilde{x}$ satisfying  $
\Phi_{J_-,n} \widetilde{x} = 0$ and $0\not=\Phi_{J_+, n}
\widetilde{x} \geq 0 $ from $\{x: Y \Phi x\geq 0\}. $ This implies that the solutions of some
existing 1-bit CS algorithms such as
\begin{eqnarray}
  &  \min \{\|x\|_1: &  Y \Phi x \geq 0 , ~\|x\|_2 =1\},    \label{Gc} \\
 & \min\{\|x\|_1:  &  Y \Phi x \geq 0 , ~\| \Phi x\|_1 = m\}    \label{Gd}
 \end{eqnarray}
   may not be consistent with the acquired
1-bit measurements. For example, let
\begin{equation} \label{EXAMPLE} \Phi = \left[
      \begin{array}{cccc}
        2 & -1 & 0 & 2 \\
        -1 & 1 & 1 & 0 \\
      \end{array}
    \right], ~ y= \left[
                  \begin{array}{c}
                    1 \\
                    -1 \\
                  \end{array}
                \right]
    .  \end{equation}
Clearly, for any scalar $\alpha >0, $  $ \widetilde{x} (\alpha) =
(\alpha,\alpha,0,0)^T \in \left\{x: Y\Phi x\geq 0\right\},$ but
$\widetilde{x}  (\alpha)  \not \in \left\{x: y=\textrm{sign}(\Phi
x)\right\} $ no matter a standard or nonstandard sign function is
used, and no matter which of the above-mentioned trivial-solution
excluders is used. Clearly, there exists a  positive number
$\alpha^*$ such that $\widetilde{x}(\alpha^*) =(\alpha^*, \alpha^*, 0, 0)^T $ is an optimal
solution to (\ref{Gc}) or (\ref{Gd}). But this optimal solution is
not consistent with $y.$

The above discussion indicates that  when $J_-
   \not = \emptyset,$    $x=0$ and $x\in {\cal N}(\Phi)$ are not contained in the set
  $ \{x: \textrm{sign}(\Phi x) =y\}.$ In this case,  we see from Lemma 2.1  that
 \begin{equation} \label{C1} \{x: \textrm{sign}(\Phi x) =y\}  \subseteq \{x: Y\Phi x\geq 0,
x\not =0 \}, \end{equation}  \begin{equation}\label{C2} \{x:
\textrm{sign}(\Phi x) =y\} \subseteq \{x: Y\Phi x\geq 0, \Phi x \not
=0 \}. \end{equation}  We now find a condition  to ensure the opposite direction of the above containing
relations.

 \vskip 0.05in

  \begin{Lem} \label{L22} Let $\textrm{sign} (\cdot)$ be the nonstandard sign function.  Let
   $\Phi\in R^{m\times n}$  and $y\in \{1, -1\}^m $
  with $J_- =\{i: y_i=-1\} \not =\emptyset$ be given.    Then    \begin{equation} \label{EEEE} \{x: Y\Phi x \geq 0, x\not =0\} \subseteq \{x:
   \textrm{sign}(\Phi x) = y \}  \end{equation}  if and only if
{\small  \begin{equation} \label{L222a} \left[\bigcup_{i \in J_-} {\cal N}(\Phi_{i, n}) \right] \cap
\left\{d: \Phi_{J_+,n}d \geq 0,  \Phi_{J_-,n}d \leq 0 \right \} =
\{0\}
\end{equation} }
where $J_+= \{i: y_i=1\}.$
\end{Lem}
 \vskip 0.05in

\emph{Proof.} Let $x$ be an arbitrary vector in the set $\{ x: Y
\Phi x\geq 0,   x \not = 0\}. $  Note that $y \in \{1, -1\}^m. $ So
$Y \Phi x\geq 0$ together with  $x\not =0$ is equivalent to
\begin{equation} \label{333}
\Phi_{J_+,n} x \geq 0, ~ \Phi_{J_-,n} x \leq 0 , ~  x \not = 0.
\end{equation}  Under the condition (\ref{L222a}), we see that for
any $x$ satisfying (\ref{333}), it must hold that $x\notin
\bigcup_{i \in J_-} {\cal N}(\Phi_{i,n})$ which implies that  $
 \Phi_{i,n}  x\not=0$ for all $ i\in J_-.$ Thus under (\ref{L222a}), the
system (\ref{333}) becomes
$  \Phi_{J_+,n} x \geq   0, \Phi_{J_-,n} x <0,   x \not = 0   $  which, by the definition of the nonstandard sign function,
 implies that  $ \textrm{sign} (\Phi x) =y. $   Thus
 (\ref{EEEE}) holds.

We now assume that the condition  (\ref{L222a}) does not hold. Then
there exists a vector $d^*\not=0$ satisfying   that
{\small \begin{equation} \label{ddd}
 d^*\in \left[\bigcup_{i \in J_-} {\cal N}(\Phi_{i,n}) \right]\cap
\left\{d: \Phi_{J_+,n}d \geq 0,  \Phi_{J_-,n}d \leq 0 \right \} .
\end{equation}}
 The fact $  d^* \in \left\{d: \Phi_{J_+,n}d \geq 0, \Phi_{J_-,n}d \leq 0
\right \}   $ implies that  $d^* \in \{x: Y \Phi x \geq 0,   x \not
= 0 \},$ and $ 0\not= d^* \in \bigcup_{i \in J_-} {\cal
N}(\Phi_{i,n})
  $
 implies that there is $ i \in J_- $ such that $\Phi_{i,n} d^* = 0.
$  By the definition of  nonstandard sign function, this implies that
$ \textrm{sign} (\Phi_{i,n} d^*)=1  \not = y_i   $ (since $y_i
=-1$ for $i\in J_-$). So $d^* \notin \{x: \textrm{sign}(\Phi x) =y
\},  $  and thus (\ref{EEEE}) does not hold.

 The above proof shows that
(\ref{EEEE}) and (\ref{L222a}) are equivalent. ~  $\Box$

\vskip 0.05in

 Replacing $x\not=0$ with $\Phi x \not=0$ and using
the same argument as above yields the next statement.

\vskip 0.05in

\begin{Lem} \label{L22a} Under the same conditions of Lemma \ref{L22},  the following statement holds:
$\{x: Y\Phi x \geq 0,   \Phi x \not =0 \} \subseteq \{x:
   \textrm{sign}(\Phi x)=y \}  $   if and only if
{\small \begin{equation}  \label{L222}   \left[\bigcup_{i \in J_-}
{\cal N}(\Phi_{i,n}) \right] \cap \left\{d: \Phi_{J_+,n}d \geq 0,
\Phi_{J_-,n}d \leq 0,  \Phi d \not =0 \right \}
  =  \emptyset.
\end{equation}}
\end{Lem} where $\emptyset$ denotes the empty set.

 \vskip 0.05in

Therefore,  we have the following result.

\vskip 0.05in

\begin{Thm} \label{Thm24} Let $\textrm{sign} (\cdot)$ be the nonstandard sign function, and let
 $\Phi\in R^{m\times n} $  and $y\in \{1, -1\}^m $ be given.
\begin{enumerate} \item[(i)] If $J_- =\emptyset,$ then   $ \{x:
    \textrm{sign}(\Phi x)= y \} = \{x: Y\Phi x \geq 0 \} .
   $
    \item[(ii)] If
   $J_-  \not =\emptyset,$  then
    $ \{x:
    \textrm{sign}(\Phi x) = y \} = \{x: Y\Phi x \geq 0, x\not =0\} $  if and only
    if (\ref{L222a}) holds.
   \item[(iii)] If
   $J_-  \not =\emptyset,$  then $ \{x:
    \textrm{sign}(\Phi x)= y \} = \{x: Y\Phi x \geq 0,   \Phi x \not =0 \}   $  if and only if
 (\ref{L222}) holds.
\end{enumerate}
\end{Thm}

The result (i)   above   is obvious. Results (ii) and (iii) follow  by
combining (\ref{C1}), (\ref{C2}) and  Lemmas 2.2 and 2.3. It is easy
to verify that the example (\ref{EXAMPLE}) does not satisfy
(\ref{L222a}) and (\ref{L222}).

 We now consider the standard sign function. In this case,  for $y=0$,  the set $\{x: Y \Phi x\geq
  0\}= R^n $  and  $ \{x: 0=\textrm{sign} (\Phi x)\} = \{x: \Phi x=0 \} = {\cal N} (\Phi) \not =
  R^n$ provided that $\Phi\not=0;$
   for $y\not =0$, we see that $ {\cal N}(\Phi) \subseteq \{x: Y \Phi x\geq
  0\}
  $  but any vector in $ {\cal N}(\Phi) $ fails to satisfy the equation $\textrm{sign}(\Phi x)
  =y. $
   Thus we have following observation:

   \vskip 0.05in

  \begin{Lem} \label {LL33}    For standard sign
  function and any nonzero $  \Phi \in R^{m\times n}, $  we have $ \{x: Y\Phi x \geq 0\}  \not=  \{x: \textrm{sign}(\Phi x) = y \}.$
\end{Lem}

 \vskip 0.05in

In general,  the set  $\{x: Y\Phi x \geq 0\}$ can be significantly
larger than $\{x: \textrm{sign}(\Phi x) = y \}.$ In what follows, we
only focus on the nontrivial case  $y\not =0. $ For a given $0\not
=y\in \{1, -1, 0\}^m$, when $J_0 =\{i: y_i=0\} \not= \emptyset,$ the
vectors in  $ {\cal N} (\Phi)$ and the vectors $ x $ satisfying $
\Phi_{J_0,n} x\not =0$   do not satisfy the constraint
$\textrm{sign}(\Phi x) =y. $  These vectors must be excluded from
$\{x: Y\Phi x\geq 0\}  $ in order to get a tighter relaxation for
the sign equation. In other words, only   vectors satisfying $\Phi
x\not=0 $ and $ \Phi_{J_0, n} x=0$, i.e., $ x\in {\cal N}
(\Phi_{J_0, n})\backslash {\cal N}( \Phi),$ should be considered.
(Note that $ {\cal N} (\Phi) \subseteq {\cal N} (\Phi_{J_0,n} )$ due
to the fact $\Phi_{J_0,n}$ being a submatrix of $\Phi.$)    Thus we
have the following result.

   \vskip 0.05in

   \begin{Thm} \label{L23} Let $\Phi\in R^{m\times n}  $ and  $0\not= y\in \{1, 0, -1\}^m$ be
   given. For the  standard sign function, the following statements hold:

(i) $\{x:   y=\textrm{sign}(\Phi x)\} \subseteq  \{x: Y\Phi x \geq
0,  \Phi_{J_0,n} x=0,  \Phi x \not =0 \} . $

(ii)  $ \{x: Y\Phi x \geq 0,\Phi_{J_0,n} x=0,  \Phi x \not =0, \}
\subseteq \{x:   \textrm{sign}(\Phi x) = y \}  $   if and only
  if
   {\small
\begin{eqnarray} \label{stand-2}  \left[\bigcup_{i \in J_+\cup J_-} {\cal N}(\Phi_{i,n})
\right]
  & \bigcap & \{d:   \Phi_{J_+,n}d \geq 0, \Phi_{J_-,n}d \leq 0, \nonumber \\
 & &     \Phi_{J_0,n} d=0,  \Phi d \not =0 \} =\emptyset.
\end{eqnarray}}
\end{Thm}

\emph{Proof.} The statement (i) follows from Lemma \ref{L21} and the
discussion before Theorem \ref{L23}. We now prove the statement (ii).
First we assume that
 (\ref{stand-2}) holds, and let $\hat{x}$ be
 an arbitrary vector in the set $ \{x: Y\Phi x \geq 0,
   \Phi_{J_0,n} x=0,  \Phi x \not =0  \} . $ Then
\begin{equation}\label{LLEE}    \Phi_{J_+,n} \hat{x} \geq 0,  ~ \Phi_{J_-,n} \hat{x} \leq 0, ~
\Phi_{J_0,n} \hat{x} =0, ~ \Phi \hat{x}\not=0.  \end{equation}    As
$y\not=0$, the set $J_+\cup J_-\not=\emptyset.$ It follows from
(\ref{stand-2}) and (\ref{LLEE})  that
$ \hat{x} \notin \bigcup_{i \in J_+\cup J_-} {\cal N}(\Phi_{i,n}),  $ which implies
that the  inequalities $\Phi_{J_+,n} \hat{x} \geq 0$ and $
\Phi_{J_-} \hat{x} \leq 0$  in (\ref{LLEE}) must hold strictly,
i.e.,
  $  \Phi_{J_+,n} \hat{x} > 0,   ~\Phi_{J_-,n} \hat{x} <
0, ~\Phi_{J_0,n} \hat{x} =0, ~ \Phi \hat{x}\not=0 ,$ and hence  $
\textrm{sign} (\Phi \hat{x}) =y .$  So
\begin{equation}\label{LLCC} \{x: Y\Phi x \geq 0,
   \Phi_{J_0,n} x=0,  \Phi x \not =0 \} \subseteq \{x: \textrm{sign}(\Phi x) =y \}.  \end{equation}

We now further prove that  if (\ref{stand-2}) does not hold, then
(\ref{LLCC}) does not hold. Indeed, assume
 that  (\ref{stand-2}) is not satisfied.  Then there
exists a vector $\hat{d}  $ satisfying $$   \Phi_{J_+,n} \hat{d}
\geq 0, ~ \Phi_{J_-,n} \hat{d} \leq 0, ~ \Phi_{J_0,n} \hat{d} =0,  ~ \Phi
\hat{d} \not=0  $$ and $$\hat{d} \in \bigcup_{i \in J_+\cup J_-}
{\cal N}(\Phi_{i,n}). $$ This implies that
$\hat{d} \in  \{x: Y \Phi x \geq 0,  \Phi_{J_0,n} x=0,  \Phi x \not
=0\} $ and that there exists
$ i\in J_+ \cup J_- $
  such that $\Phi_{i,n} \hat{d} = 0 . $ Thus $\textrm{sign} (\Phi_{i,n} \hat{d}) =0 \not= y_i$
where $y_i= 1 $ or  $ -1 $ (since $ i\in J_+ \cup J_-$). Thus
(\ref{LLCC}) does not hold.    $\Box $

\vskip 0.05in

Therefore,  under the conditions of Theorem \ref{L23}, the set $\{x:
\textrm{sign}(\Phi x) =y\} $ coincides with $ \{x: Y\Phi x \geq 0,
\Phi_{J_0,n} x=0, \Phi x \not =0 \}$ if and only if condition (\ref{stand-2})
holds. Recall that the 1-bit CS  problem  (\cite{BB2008,B2009, PV20138}) can be cast as the
$\ell_0$-minimization problem (\ref{1bitCS}), which admits the
 relaxation
 \begin{eqnarray}  & \min \{\|x\|_0:  & Y \Phi x \geq 0 , ~\|x\|_2 =1\},   \label{Ga}\\
 & \min \{\|x\|_0: &  Y \Phi x \geq 0 , ~\|\Phi x\|_1 = m\},     \label{Gb}
 \end{eqnarray}
where $m$ is not essential and  can be replaced with any positive
 constant. Replacing $\|x\|_0$ by $\|x\|_1$ immediately leads to
 (\ref{Gc}) and (\ref{Gd}) which are linear programming models.

 To guarantee that problems (\ref{Ga}) and (\ref{Gb}) are equivalent to  (\ref{1bitCS})
   and that problems (\ref{Gc}) and (\ref{Gd})  are equivalent to
 the problem
 \begin{equation} \label{G4} \min \{\|x\|_1: ~\textrm{sign}(\Phi x)=y\},
 \end{equation}
    as shown in Theorems \ref{Thm24} and \ref{L23}, the conditions (\ref{L222a}), ({\ref{L222}) or
  (\ref{stand-2}), depending on the definition of the sign function,
  must be imposed on the  matrix. These conditions   have been
   overlooked in the literature.
 If  (\ref{L222a}), ({\ref{L222}) or
  (\ref{stand-2}) is not satisfied, the feasible sets of  (\ref{Ga}), (\ref{Gb}), (\ref{Gc}) and (\ref{Gd})  are larger than
  that of (\ref{1bitCS})
   and (\ref{G4}), and thus their solutions might not satisfy the sign equation
  $\textrm{sign}(\Phi x)
  =y. $ In other words, the constructed signal through the algorithms for
  solving (\ref{Ga}), (\ref{Gb}), (\ref{Gc}) and (\ref{Gd})
  might be inconsistent
  with the acquired 1-bit measurements.


\subsection{Allowing zero in sign measurements $y$}

  The 1-bit CS
model with a nonstandard sign function does not cause any
inconvenience or difficulty when the magnitude of all components of $|\Phi x^*|$
is relatively large, in which case   $\textrm{sign} (\Phi
x^*)$ is stable in the sense that any small   perturbation of
$ \Phi  x^*$ does not affect its sign. However, when $
|\Phi  x^*| $ admits  a very small components (this case does happen in some
situations, as we point out later),  the nonstandard  sign function
might introduce certain ambiguity into the 1-bit CS model since
$\Phi x^*>0$,  $ \Phi x^*=0 $ and $0\not= \Phi x^* \geq 0 $ yield
the same
measurements  $y= (1,1, \dots , 1)^T.$ Once $y$ is acquired,
 the information concerning which of the above cases yields   $y$ in 1-bit CS models is lost.
In this situation, through sign information only,  it might  be
difficult
 to reconstruct the information of the targeted signal no matter what 1-bit CS algorithms are used.

When the magnitude of $| \Phi_{i,n} x^*|$ is very small,    errors or noises do affect the reliability of the measurements
  $y.$  The reliability   of $y$  is vital since the unknown signal is expected to
  be partially or fully reconstructed from   $y.$    Suppose that $x^*$ is the signal to recovery.
 We consider a
sensing matrix $\Phi\in R^{m\times n}$ whose rows  are  uniformly
drawn from the surface of the
$n$-dimensional unit ball $\{u\in R^n: \|u\|_2=1\}.$  Note that for any
  small positive number $\epsilon>0, $  with  positive
probability, a drawn vector lies in the region of the unit
surface
$$\{u \in R^n: \|u\|_2=1, |u^Tx^*| \leq \epsilon \}.  $$  The sensing row vector
$ \Phi_{i,n} $ drawn in this  region  yields a very small product $
\Phi_{i,n} x^* \approx 0,$ at which  $\textrm{sign} (\Phi_{i,n} x^*)
$ becomes sensitive or uncertain in the sense that  any small error
 in measuring   $\Phi_{i,n} x^*$  can totally flip its sign, leading
to an  opposite of the correct sign measurement. In this situation,
not only the acquired   information $y_i$  might be unreliable
to be used for the recover of the sign of a signal, but also the measured value $y_i= 1$ or $-1$  does not
reflect the fact $\Phi_{i,n} x^* \approx 0,$  which  indicates that $x^*$ is nearly orthogonal to the known sensing vector $\Phi_{i, n}.$ The information $\Phi_{i,n} x^* \approx 0$ is particularly useful to help locate the position of the unknown vector $x^*. $ Using only 1 or $-1$ as the sign of $\Phi_{i,n} x^*, $ however, the information  $\Phi_{i,n} x^* \approx 0 $ is completely lost
  in the 1-bit CS model.   Allowing  $y_i=0$ in this case  can
correctly reflect the relation of  $\Phi_{i,n}$ and $x^*$ when they
are nearly orthogonal. Taking into account the small magnitude of
$|\Phi_{i,n} x^*|$  and allowing $y$ to   admit zero components
  provides a practical means to avoid the aforementioned ambiguity of sign
measurements resulting from the nonstandard sign function. By using the standard sign function to distinguish the three different cases
 $\Phi x^* >0,$ $ \Phi x^* =0,$ and $ 0\not= \Phi x^* \geq 0,$   the resulting sign measurements $y$ would carry  more information of the signal, which might increase the chance for
  the sign recovery of the signal.

Thus we consider the 1-bit CS model with the standard sign function
in this paper.    In fact, the standard sign function was already
used by some authors (e.g., \cite{PV20138}) but their discussions
are based on the linear relaxation of (\ref{YYY}).

\subsection{Reformulation of 1-bit CS model}

From the above discussions, the  system (\ref{YYY}) is generally a
loose relaxation of the sign constraint of (\ref{1bitCS}). The
1-bit CS algorithms based on this relaxation might generate a
solution inconsistent with 1-bit measurements if a sensing matrix
does not satisfy the conditions specified in Theorems \ref{Thm24}
and \ref{L23}.  We  now introduce a new reformulation of the 1-bit CS model,
  which   can  ensure  that the solution of
our 1-bit CS algorithm is always consistent with the acquired 1-bit
measurements.

In the remainder of the paper, we focus on the 1-bit CS problem with
 standard sign function. For a given   $y\in \{-1, 1, 0\}^m,$  we use
$J_+,$ $ J_- $ and $J_0$ to denote  the
indices of positive, negative, and zero components of $y,$
respectively, i.e., {\small  \begin{equation} \label {JJJJ}  J_+
=\{i: y_i=1\},  J_- =\{i:~y_i=-1\},   J_0=\{i:~y_i=0\}.
\end{equation} }  Since these indices are determined by $y,$ we also write them as $J_+ (y), J_-(y)$ and $J_0 (y)$ when necessary.  By using (\ref{JJJJ}),  the   constraint $\textrm{sign}(\Phi x)=y$ can be written as
{\small \begin{equation} \label{sign-const}
 \textrm{sign} (\Phi_{J_+,n}x) =  e_{J_+},
    \textrm{sign} (\Phi_{J_-,n}x ) = - e_{J_-},
    \Phi_{J_0,n}x=0.
\end{equation} }
  Thus   the  model  (\ref{1bitCS}) with $y\in \{-1, 1, 0\}^m$
  can be stated as
 \begin{equation} \label{New-1bit} \begin{array}{cl}   \min &
\|x\|_0     \\
  \textrm{s.t.} & \textrm{sign} (\Phi_{J_+,n}x) =  e_{J_+},
   \textrm{sign} (\Phi_{J_-,n}x ) = - e_{J_-},  \\
   & \Phi_{J_0,n}x=0.
  \end{array}
  \end{equation}
 Consider the  system in $u\in R^n
$
\begin{equation}\label{var-system}
  \Phi_{J_+,n} u \geq  e_{J_+},
  ~ \Phi_{J_-,n} u \leq-  e_{J_-},
  ~ \Phi_{J_0,n} u =0.
\end{equation}
Clearly, if $x$ satisfies (\ref{sign-const}), then there exists a
positive  number $\alpha >0 $ such that $u=\alpha x$ satisfies the
system (\ref{var-system}); conversely, if $u $ satisfies the system
(\ref{var-system}), then $x=u$ satisfies the system
(\ref{sign-const}). Note that $\|x\|_0= \|\alpha x\|_0$ for any $
\alpha  \not= 0.$   Thus   (\ref{New-1bit}) can be reformulated as
the $\ell_0$-minimization problem
\begin{eqnarray}\label{l0LP}
\begin{array}{cl}
   \min & \|x\|_0 \\
            \textrm{s.t.}   & \Phi_{J_+,n}x\geq  e_{J_+},
                 ~ \Phi_{J_-,n}x\leq-  e_{J_-},
           ~ \Phi_{J_0,n}x=0.
\end{array}
\end{eqnarray}
From the relation of  (\ref{sign-const}) and (\ref{var-system}), we immediately
have the following observation.

\vskip 0.05in

\begin{Propn} \label {Prop21}  If $x^*$ is an optimal solution to the 1-bit CS model
(\ref{New-1bit}), then there exists a positive number $\alpha>0$
such that $\alpha x^*$ is an  optimal solution to the $\ell_0$-problem
(\ref{l0LP}); conversely, if $x^*$ is an optimal solution to the
$\ell_0$-problem  (\ref{l0LP}), then $x^*$ must be an optimal solution to   (\ref{New-1bit}). \end{Propn}

\vskip 0.05in

 As a result, to study the
 1-bit CS model (\ref{New-1bit}), it is sufficient to investigate the
   model (\ref{l0LP}).  This makes
 it possible to use the CS methodology
   to study the 1-bit CS problem  (\ref{New-1bit}).  Motivated by (\ref{l0LP}),
   we   consider
 the $\ell_1$-minimization
\begin{eqnarray}\label{1bit-basis}
\begin{array}{cl}
   \min & \|x\|_1 \\
            \textrm{s.t.}   & \Phi_{J_+,n}x\geq  e_{J_+},
            ~\Phi_{J_-,n}x\leq-  e_{J_-},
            ~\Phi_{J_0,n}x=0,
\end{array}
\end{eqnarray}
which can be seen as a natural decoding method for  the  1-bit CS
problems. In this paper, the problem (\ref{1bit-basis}) is referred
to as the 1-bit basis pursuit. It is worth stressing that  the
optimal solution of (\ref{1bit-basis}) is always consistent
with $y$ as indicated by Proposition \ref{Prop21}.    More
importantly, the later analysis indicates that our reformulation
makes it possible to develop a sign recovery theory for  sparse
signals from 1-bit measurements.

 For the convenience of  analysis, we define the  sets $\mathcal{A}(\cdot),$
 $\widetilde{\mathcal{A}}_+(\cdot) $ and $ \widetilde{\mathcal{A}}_-(\cdot)$ which are used
 frequently
 in this paper.  Let  $x^*\in R^n $   satisfy the constraints of
(\ref{1bit-basis}). At $x^*,$ let
\begin{equation} \label{AAA1}
\mathcal{A}(x^*)=\{i: ~(\Phi x^*)_i= 1\}\cup\{i: ~(\Phi
x^*)_i=-1 \},
\end{equation}
 \begin{equation} \label{AAA2}
\tilde{\mathcal{A}}_+(x^*)=J_+\setminus \mathcal{A}(x^*),
 ~\tilde{\mathcal{A}}_-(x^*) =J_-\setminus \mathcal{A}(x^*).
 \end{equation}
Clearly,   $\mathcal{A}(x^*)$ is the index set of active
   constraints among the inequality constraints of (\ref{1bit-basis}),  $ \tilde{\mathcal{A}}_+(x^*)$ is the index set of inactive
constraints in the first group of inequalities of (\ref{1bit-basis})
(i.e., $ \Phi_{J_+, n} x^* \geq   e_{J_+}$), and
$\tilde{\mathcal{A}}_-(x^*)$ is the index set of inactive
constraints in the second group of inequalities  of (\ref{1bit-basis}) (i.e., $\Phi_{J_-, n} x^* \leq -  e_{J_-}$).
Thus we see that
\begin{eqnarray*}
& & (\Phi x^*)_i= 1   \textrm{ for } i
\in\mathcal{A}(x^*)\cap J_+, \\
& & (\Phi x^*)_i> 1 \textrm{ for  } i\in \tilde{\mathcal{A}}_+(x^*),\\
& & (\Phi x^*)_i=-1   \textrm{ for  } i\in\mathcal{A}(x^*)\cap
J_-, \\
& & (\Phi x^*)_i<- 1 \textrm{ for
 } i\in \tilde{\mathcal{A}}_-(x^*).
\end{eqnarray*}

We also need  symbols $\pi (\cdot)$ and $\varrho (\cdot)$ defined as
follows.    Denote the  elements in $J_+ $  by $i_k \in \{1, ...,
m\}, k=1, \dots, p, $  i.e.,  $J_+=\{i_1, i_2, \dots,i_p\}$ where
$p=|J_+|.$  Without loss of generality, we let the elements be
sorted in ascending order $i_1< i_2< \cdots <i_p.$ Then we define
the bijective mapping $\pi: J_+ \to \{1, \dots, p\}$ as
\begin{equation}\label{pi} \pi( i_k) =k \textrm{ for all } k=1, \dots, p. \end{equation}     Similarly,
let $J_-=\{j_1, j_2, \dots,j_q\},$ where  $q=|J_-|,$ $j_k \in \{1, \dots, m\}$ for $k=1, \dots, q$  and $j_1< j_2<
\cdots < j_q.$  We define the bijective mapping $\varrho:  J_- \to
\{1, \dots, q\}$ as \begin{equation} \label{varrho} \varrho(j_k) =k
~\textrm{ for all } k=1, \dots, q.
\end{equation}

By introducing   variables $\alpha\in R^{|J_+|}_+$ and $\beta\in
R^{|J_-|}_+$, the  problem (\ref{1bit-basis})
 can be written as
\begin{eqnarray}\label{8888}
  & \min &      \|x\|_1,  \nonumber \\
        & \textrm{ s.t.}& \Phi_{J_+,n} x -\alpha=  e_{J_+}, \nonumber\\
        &      & \Phi_{J_-,n} x +\beta=-  e_{J_-}, \\
        &      & \Phi_{J_0,n} x=0, \nonumber \\
        &      & \alpha\geq0,~\beta\geq0.\nonumber
\end{eqnarray}
Note that for any optimal solution $(x^*, \alpha^*, \beta^*)  $ of
(\ref{8888}),  we have $\alpha^*=\Phi_{J_+,n} x^* -   e_{J_+}
$ and $ \beta^* =-  e_{J_-}- \Phi_{J_-,n} x^*.  $
 Using  (\ref{AAA1})--(\ref{varrho}), we immediately have the following
observation.

\vskip 0.05in

\begin{Lem} \label{LemlepsLP1} (i) For any optimal solution  $(x^*, \alpha^*, \beta^*)$ to the problem (\ref{8888}),
we have
\begin{equation} \label{Lemma31}
\left\{\begin{array}{ll}
  \alpha^*_{\pi(i)}=0, & \text{for } i\in \mathcal{A}(x^*)\cap J_+, \\
  \alpha^*_{\pi(i)}= (\Phi x^*)_i -1 >0, & \text{for } i\in \tilde{\mathcal{A}}_+(x^*), \\
 \beta^*_{\varrho(i)} =0, & \text{for } i\in \mathcal{A}(x^*)\cap J_-, \\
 \beta^*_{\varrho(i)}=-1 -(\Phi x^*)_i>0, & \text{for } i\in \tilde{\mathcal{A}}_-(x^*).
\end{array} \right.
\end{equation}
(ii)   $x^*$ is the unique optimal solution to   the 1-bit basis pursuit
(\ref{1bit-basis}) if and only if $ (x^*,\alpha^*,\beta^*)$ is the
unique  optimal solution to the problem (\ref{8888}),  where  $(\alpha^*,
\beta^*)$ is determined by (\ref{Lemma31}).

\end{Lem}

\subsection{Recovery criteria}

When   $y=\textrm{sign} (\Phi x^*) \in \{1, -1\}^m,$    any small
perturbation $x^*+  u $  is  also
 consistent with $y. $ When $y\in \{1, -1, 0\}^m ,$   any small perturbation
  $x^*+  u $ with $ u\in {\cal N} (\Phi_{J_0, n})   $ is  also consistent with $y.$
  Thus a 1-bit CS problem generally has infinitely many solutions and the
  sparsest solution of a sign equation is also not unique in general. Since the amplitude of signals is not available,
   the recovery criteria in 1-bit CS scenarios can be
    sign recovery, support recovery or others, depending on signal environments. The exact sign recovery
  of a signal means  that the found solution $\widetilde{x}$ by an algorithm satisfies
  $$\textrm{sign} (\widetilde{x})= \textrm{sign}(x^*). $$    The support recovery, i.e., the found
  solution $\widetilde{x}$ satisfying
 $ \textrm{supp} (\widetilde{x})= \textrm{supp}(x^*)$ is  a relaxed version of the sign recovery.
It is worth mentioning that the following criterion $$ \left\| \frac{x}{\|x\|_2}  - \frac{x^*}{\|x^*\|_2}  \right\| \leq
\varepsilon $$ has been  widely used in the 1-bit CS literature,
  where $\varepsilon>0 $ is a certain small number.

In the remainder of the paper, we work toward developing some necessary and sufficient conditions for the
  exact  recovery of the sign of sparse signals from  1-bit measurements.

\section{Nonuniform sign recovery}

   We assume that the  measurements $y=\textrm{sign} (\Phi x^*) $ is
available. From this information,  we use the 1-bit basis pursuit
(\ref{1bit-basis})  to recover the sign of  $x^*.$  We ask when the
optimal solution of (\ref{1bit-basis}) admits the same sign of
$x^*.$  The recovery of the sign of an individual sparse signal is
referred to as the nonuniform sign recovery. In this section, we
develop certain necessary and sufficient conditions for the
nonuniform sign  recovery  from the perspective of the range space
property of a transposed sensing matrix.

Assume that $y\in \{1, -1, 0\}^m$ is given and $(J_+, J_-, J_0)$ is specified as (\ref{JJJJ}).
 We first introduce the concept of the RRSP.

\vskip 0.05in

\begin{Def} [RRSP of $\Phi^T$ at $x^*$] \label{DefRRSP} Let  $x^*\in R^n$ satisfy $y=\textrm{sign} (\Phi x^*).$ We say that
 $\Phi^T$ satisfies  the restricted range space property (RRSP) at
 $x^*$
  if there exist  vectors $\eta\in
\mathcal{R}(\Phi^T)$  and $w\in \mathcal{F}(x^*)$  such that $\eta=\Phi^T w$  and
$$\eta_i=1   \text{ for } x^*_i>0,
  \eta_i=-1  \text{ for } x^*_i<0,
  |\eta_i|<1 \text{ for } x^*_i=0,$$
where  $ \mathcal{F}(x^*) $ is the set defined as
 \begin{eqnarray} \label{FFFF}
 &   \mathcal{F}(x^*) & =  \{w \in R^m:  w_i >  0\textrm{ for } i\in {\cal A}(x^*) \cap
 J_+, \nonumber \\  &   & ~~~~ w_i < 0 \textrm{ for } i\in {\cal A}(x^*) \cap J_-, \\
  &   & ~~~~  w_i=0 \textrm{ for } i\in \tilde{\mathcal{A}}_+(x^*) \cup \tilde{\mathcal{A}}_-(x^*)\}. \nonumber
 \end{eqnarray}
\end{Def}

The RRSP of $\Phi^T$ at $x^*$ is a natural condition for the
uniqueness of optimal solutions to the 1-bit basis pursuit
(\ref{1bit-basis}), as shown by the following theorem.

\vskip 0.05in

\begin{Thm}[Necessary and sufficient condition]\label{Ness-Suff}
 $x^*$ is the unique optimal solution to the 1-bit basis pursuit (\ref{1bit-basis})  if and only
if the RRSP of $\Phi^T $ at $x^*$  holds and the matrix  \begin{eqnarray}\label{FRPmatrix}
H (x^*) =\left[
         \begin{array}{cc}
           \Phi_{\mathcal{A}(x^*)\bigcap J_+, S_+} & \Phi_{\mathcal{A}(x^*)\bigcap J_+, S_-} \\
           \Phi_{\mathcal{A}(x^*)\bigcap J_-, S_+} & \Phi_{\mathcal{A}(x^*)\bigcap J_-, S_-} \\
           \Phi_{J_0,S_+} & \Phi_{J_0,S_-}
                  \end{array}
       \right]
\end{eqnarray}  has a full-column rank, where $S_+=\{i: x^*_i>0\} $ and $ S_-=\{i: x^*_i<0\}.$
 \end{Thm}

\vskip 0.05in

The proof of Theorem 3.2 requiring some fundamental facts for linear
programs is given in Section V. 
The uniqueness of solutions to a decoding method like
(\ref{1bit-basis}) is an important
 property required in signal reconstruction.  As
indicated in \cite{F2004, P07, FR2013, YBZ2013}, the
uniqueness conditions  often lead to certain criteria for the
nonuniform and uniform recovery of sparse signals. Later, we will
see that  Theorem \ref{Ness-Suff},  together with the matrix
properties N-RRSP and S-RRSP of order $k$ that will be introduced in
this and next sections, provides a fundamental basis to develop a
sign recovery theory for sparse signals from   1-bit  measurements.
Let us begin with the following lemma.

\vskip 0.05in

\begin{Lem}\label{Lemma41} Let $x^*$ be a sparsest solution of the
$\ell_0$-problem (\ref{l0LP}) and let $S_+ $ and $
S_-  $  be defined as  in Theorem \ref{Ness-Suff}. Then
\begin{equation}\label{HHTT} \widetilde{H}(x^*) = \left[\begin{array}{cc}
                 \Phi_{\mathcal{A}(x^*)\cap J_+,S_+} & \Phi_{\mathcal{A}(x^*)\cap J_+,S_- } \\
                 \Phi_{\mathcal{A}(x^*)\cap J_-,S_+} & \Phi_{\mathcal{A}(x^*)\cap J_-,S_- } \\
                 \Phi_{J_0,S_+} & \Phi_{J_0,S_-} \\
                 \Phi_{\tilde{\mathcal{A}}_+(x^*),S_+} & \Phi_{\tilde{\mathcal{A}}_+(x^*),S_- } \\
                 \Phi_{\tilde{\mathcal{A}}_-(x^*),S_+} & \Phi_{\tilde{\mathcal{A}}_-(x^*),S_- }
               \end{array}\right]
\end{equation}  has a full-column rank.  Furthermore, at any sparsest solution $x^*$  of
(\ref{l0LP}), which admits the maximum cardinality  $ |{\cal
A}(x^*)| =\max\{|{\cal A}(x)|: x\in F^*\}, $ where $F^*$ is the set of optimal
solutions
 of (\ref{l0LP}),
  $ H(x^*)    $  given by (\ref {FRPmatrix})   has a
full-column rank.
\end{Lem}

\vskip 0.05in

{\it Proof.}  Note that $x^*$ is a sparsest solution to the
  system
\begin{equation} \label{sys0}
  \Phi_{J_+,n} x^* \geq   e_{J_+},
 ~ \Phi_{J_-,n} x^* \leq -  e_{J_-},
 ~  \Phi_{J_0,n} x^*=0.
\end{equation}
%
Including $\alpha^*$ and $\beta^*,$ given by (\ref{Lemma31}), into
  (\ref{sys0})  leads to
{\small \begin{equation}\label{sys4}
  \Phi_{J_+,n}x^*-\alpha^*=  e_{J_+},
  ~\Phi_{J_-,n}x^*+\beta^*=-  e_{J_-},
  ~ \Phi_{J_0,n}x^*=0.
\end{equation} }
Eliminating the zero components of $x^* $ from (\ref{sys4}) leads to
\begin{eqnarray}\label{sys5}
\left\{\begin{array}{l}
\Phi_{ J_+,S_+} x^*_{S_+}+\Phi_{ J_+,S_- }x^*_{S_-} -\alpha^*
=  e_{ J_+},\\
\Phi_{  J_-,S_+} x^*_{S_+}+\Phi_{  J_-,S_- }x^*_{S_-} +\beta^*
=-  e_{  J_-},\\
\Phi_{J_0,S_+} x^*_{S_+}+\Phi_{J_0,S_-} x^*_{S_-}=0.
\end{array} \right.
\end{eqnarray}
Since $x^*$ is a sparsest solution of (\ref{l0LP}), it is not very
difficult to see that the coefficient matrix
 $$ \widehat{ H }  =\left[\begin{array}{cc}
                 \Phi_{J_+,S_+} & \Phi_{ J_+,S_- } \\
                 \Phi_{J_-,S_+} & \Phi_{ J_-,S_- } \\
                 \Phi_{J_0,S_+} & \Phi_{J_0,S_-}
               \end{array}\right]
$$
has  a full-column rank, since otherwise at least one column of $
 \widehat{ H } $ can be linearly represented by its other columns,   the system
(\ref{sys5}), which is equivalent to  (\ref{sys0}), has a solution
  sparser than $x^*. $    From (\ref{AAA1}) and
(\ref{AAA2}), we see that
{\small \begin{equation} \label{XXXc} J_+ =
 (\mathcal{A}(x^*)\cap J_+)\cup \widetilde{\mathcal{A}}_+(x^*), ~ J_-
=  (\mathcal{A}(x^*)\cap J_-) \cup\widetilde{\mathcal{A}}_-(x^*).
\end{equation}    }
Performing row
permutations on $ \widehat{ H }, $ if necessary, yields
$\widetilde{H}(x^*)$ given as
    (\ref{HHTT}).
Since row permutations do not affect the column rank of $
\widehat{H} ,$ $ \widetilde{H}(x^*)$ must have a full-column rank.

We now show that  $H(x^*)$ has a full-column rank if ${\cal A}(x^*)$
  admits the  maximum cardinality   in the sense that
   $ |{\cal A}(x^*)|  =\max\{|{\cal A} (x)|: x \in F^*\}, $
   where $F^*$ is the set of optimal solutions  of (\ref{l0LP}).  We prove this by contradiction.
    Assume that the columns of $H(x^*)$ are
linearly dependent. Then there is a nonzero vector  $ d= (u,v)  \in
R^{|S_+|}\times R^{|S_-|}  $ such that $$H(x^*) d= H(x^*) \left[
                                   \begin{array}{c}
                                     u \\
                                     v \\
                                   \end{array}
                                 \right] =0.$$ Since $d  \not =0$  and  $\widetilde{H}(x^*),$ given by
                                 (\ref{HHTT}),
                                 has a full-column rank, we see that
               \begin{equation} \label{HHdd}
               \left[\begin{array}{cc}
                 \Phi_{\tilde{\mathcal{A}}_+(x^*),S_+} & \Phi_{\tilde{\mathcal{A}}_+(x^*),S_- } \\
                 \Phi_{\tilde{\mathcal{A}}_-(x^*),S_+} & \Phi_{\tilde{\mathcal{A}}_-(x^*),S_- }
               \end{array}\right] \left[
                                   \begin{array}{c}
                                     u \\
                                     v \\
                                   \end{array}
                                 \right]  \not =0.
               \end{equation}
Let $x(\lambda) $ be the vector with components  $ x(\lambda)_{S_+}
= x^*_{S_+} + \lambda u,$ $ ~x(\lambda)_{S_-}  = x^*_{S_-} + \lambda v
$ and  $ x(\lambda)_i=0 \textrm{  for all  }i \notin S_+\cup S_-,$
where $\lambda\in R. $ Clearly, we have  $\textrm{supp}(x(\lambda))
\subseteq \textrm{supp} (x^*)$ for any $\lambda\in R. $
 By (\ref{Lemma31}) and (\ref{XXXc}),  the system (\ref{sys5}) is equivalent to
    \begin{equation}\label{system1}
\left\{\begin{array}{l} \Phi_{\mathcal{A}(x^*)\cap J_+,S_+}
x^*_{S_+}+\Phi_{\mathcal{A}(x^*)\cap J_+,S_- }x^*_{S_-}
=  e_{\mathcal{A}(x^*)\cap J_+},\\
\Phi_{\mathcal{A}(x^*)\cap J_-,S_+}
x^*_{S_+}+\Phi_{\mathcal{A}(x^*)\cap J_-,S_- }x^*_{S_-}
=-  e_{\mathcal{A}(x^*)\cap J_-},\\
\Phi_{J_0,S_+} x^*_{S_+}+\Phi_{J_0,S_-} x^*_{S_-}=0,\\
\Phi_{\tilde{\mathcal{A}}_+(x^*),S_+}
x^*_{S_+}+\Phi_{\tilde{\mathcal{A}}_+(x^*),S_- }x^*_{S_-}
  >  e_{\tilde{\mathcal{A}}_+(x^*)},\\
\Phi_{\tilde{\mathcal{A}}_-(x^*),S_+}
x^*_{S_+}+\Phi_{\tilde{\mathcal{A}}_-(x^*),S_- }x^*_{S_-}< -
e_{\tilde{\mathcal{A}}_-(x^*)},
\end{array} \right.
\end{equation}
From the above system and the definition of $x(\lambda),$  we see
that for any sufficiently small $|\lambda| \not=0, $ the vector   $(
x(\lambda)_{S_+},$ $
                x(\lambda)_{S_-})$ satisfies the system
{\small \begin{eqnarray} &  H(x^*) \left[
             \begin{array}{c}
               x(\lambda)_{S_+}
                \\
                 x(\lambda)_{S_-}
             \end{array}
           \right] = \left[
                       \begin{array}{c}
                          e_{\mathcal{A}(x^*)\cap J_+} \\
                          -e_{\mathcal{A}(x^*)\cap J_-} \\
                         0 \\
                       \end{array}
                     \right],   &  \label {H1} \\
 &    \left[\Phi_{\widetilde{\mathcal{A}}_+(x^*),S_+}, \Phi_{\widetilde{\mathcal{A}}_+(x^*),S_-}\right] \left[
             \begin{array}{c}
               x(\lambda)_{S_+}
                \\
                 x(\lambda)_{S_-}
             \end{array}
           \right]   > e_{\widetilde{\mathcal{A}}_+(x^*)}, ~~&  \label {H2} \\
&   \left[\Phi_{\widetilde{\mathcal{A}}_- (x^*),S_+}, \Phi_{\widetilde{\mathcal{A}}_- (x^*),S_-}\right] \left[
             \begin{array}{c}
               x(\lambda)_{S_+}
                \\
                 x(\lambda)_{S_-}
             \end{array}
           \right]    <- e_{\widetilde{\mathcal{A}}_-(x^*)}. ~~  &  \label{H3}
 \end{eqnarray} }
Equality (\ref{H1}) actually holds for any $\lambda \in R^n. $
 Starting from $\lambda =0,$ we continuously increase the value of $|\lambda|$. In this process, if one of the components of the vector $( x(\lambda)_{S_+},
                x(\lambda)_{S_-})$ satisfying (\ref{H1})--(\ref{H3}) becomes zero, then a
                sparser solution than $x^*$ is found, leading to a contradiction. Thus
                without loss of generality, we assume that $ \textrm{supp} (x(\lambda)) = \textrm{supp}(x^*)$
                 is maintained when $|\lambda|$ is continuously increased. It follows from
                 (\ref{HHdd}) that
                 there exists $\lambda^* \not=0 $   such that $(x (\lambda^*)_{S_+}, x( \lambda^*)_{S_-})$ satisfies  (\ref{H1})--(\ref{H3}) and at this vector,    one of the inactive constraints
                 in (\ref{H2}) and (\ref{H3})   becomes active. Therefore $|{\cal A} (x(\lambda^*))| > |{\cal A} (x^*)|.$
                  This contradicts the fact $|{\cal A} (x^*)|$ has the maximal cardinality amongst
                  the sparsest solutions.  Thus we conclude that  $H(x^*)$ must have a full-column rank. ~~ $ \Box $

\vskip 0.05in

From Lemma \ref{Lemma41}, we see that the full-rank property of
(\ref {FRPmatrix}) can be guaranteed if $x^*$ is a sparsest solution
consistent with 1-bit measurements
 and $ | \mathcal{A} (x^*)|$ is maximal.   Thus by Theorem \ref{Ness-Suff}, the central
 condition  for $x^*$ to be the unique optimal solution  to  (\ref{1bit-basis}) is the RRSP
  described in Definition \ref{DefRRSP}.  From the above discussions, we
  obtain the following  connection
between  1-bit CS  and 1-bit basis pursuit.

\vskip 0.05in

\begin{Thm}\label{Thm-equ}
 (i) Suppose that $x^*$ is an optimal solution to the $\ell_0$-problem (\ref{l0LP})
  with maximal $| \mathcal{A} (x^*)|.$   Then $x^* $ is the unique optimal solution to  (\ref{1bit-basis})
   if and only if the RRSP of $\Phi^T$
  at $x^* $  holds.
  (ii)  Suppose that $x^*$ is an optimal solution  to the problem (\ref{New-1bit}) or (\ref{l0LP}).
  Then the sign of $x^*$ coincides with the sign of the unique solution of
   (\ref{1bit-basis}) if and only if  there exists a weight $z\in R^n $ satisfying $z_i> 0$ for $i\in \textrm{supp}
   (x^*)$ and $z_i =0$ for $i\notin \textrm{supp} (x^*)$ such that $Zx^*,$  where $Z=\textrm{diag} (z),$
   is feasible to (\ref{1bit-basis}) and $H(Zx^*)$ has a full-column rank and the RRSP of $\Phi^T$ at $Zx^*$ holds.
\end{Thm}

\vskip 0.05in

\emph{Proof. } Result (i) follows directly from Lemma \ref{Lemma41} and Theorem \ref{Ness-Suff}. We now prove
  result (ii).   If the sign of $x^*$  coincides with the sign of the unique optimal solution $\widetilde{x}$ of
  (\ref{1bit-basis}), then $\widetilde{x}$  can be written as $ \widetilde{x}= Zx^*$ for a certain weight
   satisfying that  $z_i> 0$ for $i\in \textrm{supp} (x^*)$ and $z_i =0$ for $i\notin \textrm{supp} (x^*).$
   It follows from Theorem \ref{Ness-Suff}  that $H(Zx^*)$ has a full-column rank and the RRSP of
    $\Phi^T$ at $Zx^*$ holds.
Conversely, if   there exists a weight $z \in R^n $ satisfying $z_i>
0$ for $i\in \textrm{supp} (x^*)$ and $z_i =0$ for $i\notin
\textrm{supp} (x^*)$ such that $ \widetilde{x} = Zx^*,$ where
$Z=\textrm{diag} (z),$ is feasible to (\ref{1bit-basis}) and
$H(Zx^*)$ has a full-column rank and the RRSP of $\Phi^T$ at $Zx^*$
holds, then by Theorem \ref{Ness-Suff} again    $\widetilde{x}=Zx^*$
is the unique optimal solution to (\ref{1bit-basis}). Clearly, by
the definition of $Z,$ we have $\textrm{sign} (\widetilde{x} )
 =\textrm{sign} (Zx^*)=\textrm{sign} (x^*). $   ~$\Box$

\vskip 0.05in

The above result  provides some insight into the nonuniform recovery
of the sign of   an individual sparse signal via the 1-bit
measurements and 1-bit basis pursuit. This result indicates that
central to the sign recovery of $x^*$  is the RRSP of $\Phi^T$  at
$x^*.$ However, this property is defined at  $x^*,$ which is unknown
in advance. Thus we need to further strengthen this concept in order
to develop certain   recovery conditions independent of the specific
signal $x^*.$ To this purpose, we introduce the notion of \emph{N-
and  S-RRSP of order $k$ with respect to 1-bit measurements,} which
turns out to be  a necessary condition and a sufficient condition,
respectively, for the nonuniform sign recovery.

For given measurements $y\in \{1, -1, 0\}^m,  $  let   $P(y) $ denote the set
of all possible partitions of the support of signals consistent with
  $y$:
$$P(y)= \{ (S_+ (x), S_- (x)):   y=\textrm{sign} (\Phi x) \}$$
where $S_+(x)= \{i: x_i >0\} $ and $S_-(x)= \{i: x_i < 0\} .$

\vskip 0.05in

 \begin{Def}[N-RRSP of order $k$ with respect to $y$]\label{NRRSP-y}  The matrix $\Phi^T$
is said to satisfy the necessary restricted range space property
(N-RRSP) of order $k$ with respect to $y$  if there exist a   pair
$(S_+,
  S_-) \in P(y) $  with $|S_+ \cup S_-|\leq k$  and
a pair $(T_1, T_2)$ with
  $T_1 \subseteq J_+$, $T_2 \subseteq J_-,$ $ T_1\cup T_2 \not =J_+\cup J_-  $ and
 $  \left[\begin{array}{c}
     \Phi_{J_+\backslash T_1,S} \\
     \Phi_{J_-\backslash T_2,S} \\
     \Phi_{J_0,S}
   \end{array}\right],  $ where $S =S_+ \cup S_-,$  having a full-column rank
   such that there is a vector $\eta \in {\cal R} (\Phi^T) $ satisfying the following properties:
   \begin{enumerate}
   \item[(i)] $
  \eta_i=1  \text{ for }i\in S_+, ~
  \eta_i=-1 \text{ for }i\in S_-, ~ |\eta_i|<1 \text{ otherwise};
 $
 \item[(ii)]
   $\eta=\Phi^T w$  for some $w\in \mathcal{F}(T_1,T_2), $ where
\begin{eqnarray}\label{F4} & \mathcal{F}(T_1,T_2) =  \{w\in R^m : &  w_{J_+ \backslash T_1} >  0, w_{J_-
 \backslash T_2}< 0, \nonumber \\
 & & w_ {T_1 \cup T_2}=0 \}.
\end{eqnarray}
\end{enumerate}
\end{Def}

The above matrix property turns out to be a necessary condition for
the nonuniform recovery of the sign of a $k$-sparse signal, as shown
by the next theorem. \vskip 0.05in

\begin{Thm}\label{Main36} Let $x^*$ be an unknown $k$-sparse
signal (i.e., $\|x^*\|_0\leq k$) and
 assume that the measurements $y=\textrm{sign}(\Phi x^*)  $ are known. If
 the 1-bit basis pursuit (\ref{1bit-basis}) admits a unique optimal solution $\widetilde{x} $ satisfying $ \textrm{sign} (\widetilde{x}) =\textrm{sign}(x^*)$ (i.e., the sign  of $x^*$ can be exactly recovered
 by (\ref{1bit-basis})), then $\Phi^T$ has
the N-RRSP of order $k$ with respect to $y$.
\end{Thm}

 \vskip 0.05in

\emph{Proof.} Suppose that the measurements $y =\textrm{sign} (\Phi
x^*)$ are  given,   where $x^*$  is an unknown $k$-sparse signal. By
the definition of $P(y),$ we see that $ (S_+(x^*), S_- (x^*)) \in
P(y). $ Denote by $S=S_+(x^*)\cup S_-(x^*).$  Suppose that
(\ref{1bit-basis}) has a unique optimal solution $\widetilde{x} $
satisfying $ \textrm{sign} (\widetilde{x}) =\textrm{sign}(x^*), $
which implies that $ (S_+(\widetilde{x}), S_-(\widetilde{x})) =
(S_+(x^*), S_-(x^*)). $ By Theorem \ref{Ness-Suff},  the uniqueness
of $ \widetilde{x} $ implies that the RRSP of $\Phi^T$  at
$\widetilde{x} $ holds and   $H(\widetilde{x}) $ has a
full-column rank. Let
\begin{equation} \label{T1T2} T_1= \widetilde{{\cal A}}_+(\widetilde{x})=J_+\setminus {\cal A}(\widetilde{x}) ,  T_2= \widetilde{{\cal A}}_-(\widetilde{x}) = J_-\setminus
{\cal A}(\widetilde{x}) . \end{equation} Note that at any optimal solution of (\ref{1bit-basis}), at least one of the inequality constraints of (\ref{1bit-basis}) must be active. Thus $  {\cal A}(\widetilde{x}) \not =\emptyset,$ which implies that  $ T_1\cup T_2 \not =J_+\cup J_-.  $
We also note that $ J_+\setminus T_1
=J_+\cap{\cal A}(\widetilde{x})   $ and $ J_-\backslash T_2 = J_-
\cap   {\cal A}(\widetilde{x}).$   Hence the matrix $ \left[\begin{array}{c}
     \Phi_{J_+\backslash T_1,S} \\
     \Phi_{J_-\backslash T_2,S} \\
     \Phi_{J_0,S}
   \end{array}\right],$  coinciding with $H(\widetilde{x}), $    has a full-column rank.
   The RRSP of $\Phi^T$ at $\widetilde{x}$ implies that properties (i)
   and (ii) of Definition \ref{NRRSP-y} are satisfied with
   $(S_+,S_-)=(S_+(\widetilde{x}), S_-(\widetilde{x})) = (S_+(x^*), S_-(x^*))$ and
   $(T_1, T_2)$ being given as (\ref{T1T2}).
    This implies that the N-RRSP of order $k$ with respect to $y$ must hold. ~ $\Box$

\vskip 0.05in

A slight enhancement of the N-RRSP property by varying the choices of
$(S_+, S_-)$ and $(T_1, T_2)$, we obtain the next  property
which turns out to be a sufficient condition for the  exact recovery of the sign
of a $k$-sparse signal.

\vskip 0.05in

 \begin{Def}[S-RRSP of order $k$ with respect to $y$]\label{SRRSP-y}  The matrix $\Phi^T$
is said to satisfy the sufficient restricted range space property
(S-RRSP) of order $k$ with respect to $y$  if for any    $(S_+
, S_-) \in P(y) $   with $|S_+ \cup S_- |\leq k$, there exists a pair $(T_1, T_2)$ such that $T_1
\subseteq J_+$, $T_2 \subseteq J_- , $   $  T_1\cup T_2 \not =J_+\cup J_-  $and $
\left[\begin{array}{c}
     \Phi_{J_+\backslash T_1,S} \\
     \Phi_{J_-\backslash T_2,S} \\
     \Phi_{J_0,S}
   \end{array}\right], $  where $S =S_+ \cup S_-,$ has a full-column rank,  and for any such a pair $(T_1, T_2),$
    there is a vector $\eta \in
{\cal R}(\Phi^T)$ satisfying the following properties:
\begin{enumerate} \item [(i)] $
  \eta_i=1  \text{ for }i\in S_+,
 ~  \eta_i=-1 \text{ for }i\in S_-,
   ~ |\eta_i|<1 \text{ otherwise};
$
\item[(ii)]   $\eta=\Phi^T w$  for some $w\in \mathcal{F}(T_1,T_2) $ defined
by (\ref{F4}).
\end{enumerate}
\end{Def}

\vskip 0.05in

Note that when $\left[\begin{array}{c}
     \Phi_{J_+\backslash T_1,S} \\
     \Phi_{J_-\backslash T_2,S} \\
     \Phi_{J_0,S}
   \end{array}\right]$ has a full-column rank, so does $\Phi_{m, S}.$ Thus we have the next lemma.

\vskip 0.05in

\begin{Lem}\label{KcolumnRSP}  If $\Phi^T$ satisfies the S-RRSP of order $k$ with
respect to  $y, $  then for any $(S_+, S_-)\in P(y) $ with $|S_+\cup
S_-|\leq k,$  $\Phi_{m, S} $ must have a full-column rank, where
$S=S_+\cup S_-.$

\end{Lem}

\vskip 0.05in

For a given $y , $ the equation $y=\textrm{sign}(\Phi x)$ might
possess infinitely many solutions. We now prove that if   $x^*$ is a
sparsest solution to this equation, then its sign can be exactly
recovered by (\ref{1bit-basis}) if    $\Phi^T$ has the S-RRSP of
order $k$ with respect to $y.$

\vskip 0.05in

\begin{Thm}\label{Main38}
 Let   measurements $y\in \{-1, 1,0\}^m$ be given and assume that $\Phi^T$ has
the S-RRSP of order $k$ with respect to   $y$. Then
 the 1-bit basis pursuit (\ref{1bit-basis}) admits a unique optimal solution $x' $
 satisfying $ \textrm{supp} (x') \subseteq \textrm{supp}(x^*) $ for any
$k$-sparse signal $x^*$  consistent with the measurements $ y,$
i.e.,  $y= \textrm{sign}(\Phi x^*). $ Furthermore, if $x^*$ is a
sparsest signal consistent with $y,$
 then $ \textrm{sign} (x') =\textrm{sign}(x^*),$ and thus the sign  of $x^*$ can be exactly recovered
 by  (\ref{1bit-basis}).
\end{Thm}

\vskip 0.05in

{\it Proof.}  Let $x^*$ be a  $k$-sparse signal consistent with $y$,
i.e., $\textrm{sign}(\Phi x^*) =y. $  Denote by
$S_+=\{i: x^*_i>0\}$, ~ $S_-=\{i: x^*_i<0\}$ and
$S=\textrm{supp}(x^*)= S_+\cup S_-.$ Clearly, $(S_+, S_-) \in P(y)$
and $|S_+\cup S_-|\leq k.$   Consistency implies that $ (\Phi x^*)_i
>0\textrm{ for all } i\in J_+,   (\Phi x^*)_i <
0\textrm{ for all } i\in J_- $ and $  (\Phi x^*)_i = 0\textrm{ for
all } i\in J_0.$ This implies that  there is a scalar $\alpha>0$
such that $ \alpha (\Phi x^*)_i \geq 1 \textrm{ for all } i\in J_+ $
 and $  \alpha (\Phi x^*)_i \leq -1  \textrm{ for all } i\in J_-. $
Thus $\alpha x^*$ is feasible to (\ref{1bit-basis}), i.e.,
\begin{eqnarray}
& &\Phi_{J_+,n} (\alpha x^*) \geq   e_{J_+},\label{eqn13a}\\
& &\Phi_{J_-,n} (\alpha x^*) \leq -   e_{J_-},\label{eqn14a}\\
& &\Phi_{J_0,n} (\alpha x^*) =0. \label{eqn15a}
\end{eqnarray}
We see that
$\alpha \geq \frac{1}{ (\Phi x^*)_i}\textrm{
for }i\in J_+ $ and $ \alpha \geq \frac{1}{-(\Phi x^*)_i}\textrm{ for
}i\in J_-. $  Let $\alpha^* $ be the smallest $\alpha$ satisfying these
 inequalities, i.e.
{\small $$ \alpha^* =  \max \left\{ \max_{i\in J_+} \frac{1}{ (\Phi x^*)_i},   \max_{i\in J_-}
 \frac{1}{-(\Phi x^*)_i} \right\} =\max_{i\in J_+\cup J_-}  \frac{1}{ |(\Phi x^*)_i|}.$$ }
By the   choice of $\alpha^*$,   at $\alpha^* x^* $   one of the
inequalities in (\ref{eqn13a}) and (\ref{eqn14a}) becomes an
equality. Let $T'_0$ and $T''_0$ be the set of indices for
 active constraints  in (\ref{eqn13a}) and
(\ref{eqn14a}),   i.e.,
{\small  $$  T'_0  =
\left\{i \in J_+ :    \Phi  (\alpha^* x^*) _i =1  \right\},
    T''_0  =  \left\{i \in J_- :     \Phi  (\alpha^* x^*) _i =-1  \right\} $$ }
If the null space $ {\cal N} (\left[
                          \begin{array}{c}
                             \Phi_{ T'_0, S} \\
                            \Phi_{ T''_0, S} \\
                            \Phi_{J_0, S}  \\
                          \end{array}
                        \right] ) \not=\{0\},$
then let $d \not=0 $ be a vector in this null space. It follows from
Lemma \ref{KcolumnRSP} that  $\Phi_{m,S}$ has a full-column rank.
This implies that \begin{equation}\label{HD} \left[
\begin{array}{c}
 \Phi_{J_+\backslash T'_0, S} \\
  \Phi_{ J_-\backslash T''_0, S}
  \end{array}
  \right] d \not =0.\end{equation}
 Consider the vector
$x(\lambda)$ with components $ x(\lambda)_S = \alpha^* x^*_S +
\lambda d$ and $x(\lambda)_i =0 $ for $i\notin S,$ where $\lambda
\in R.$  By the choice of $d$, we see that $ \textrm{supp}(
x(\lambda))\subseteq \textrm{supp}(x^*) $ for any $\lambda \in R.$
For all sufficiently small $ |\lambda|, $ the vector $ x(\lambda)  $
is feasible to the problem (\ref{1bit-basis}) and the  active
constraints at $\alpha^* x^* $ in (\ref{eqn13a}) and (\ref{eqn14a})
are still active at $ x(\lambda) $  and the inactive constraints at
$\alpha^* x^* $ are still inactive at $ x(\lambda). $ Due to
(\ref{HD}), if letting $|\lambda|$ continuously vary  from zero to a
positive number, there exists  $ \lambda^* \not =0 $ such that
$x(\lambda^*) $ is still feasible to (\ref{1bit-basis}) and one of
the above-mentioned inactive constraints  becomes active at
$x(\lambda^*).$  Let $x' = x(\lambda^*)$ and
 $$T'  =  \left\{i \in J_+ :    (\Phi  x' )_i =1
 \right\},
   T''  = \left\{i \in J_- :    (\Phi  x')_i =-1  \right\}. $$
 By the construction of $x',$ we see that $ T'_0 \subseteq T' $ and $ T''_0 \subseteq T''.$  So we obtain an augmented
set of active constraints at $x'.$

Now replace the role of $\alpha^* x^*$ by $ x'$ and repeat the above
process.
 If $ {\cal N}  (\left[
                          \begin{array}{c}
                             \Phi_{ T', S} \\
                            \Phi_{ T'', S} \\
                            \Phi_{J_0, S}  \\
                          \end{array}
                        \right])  \not=\{0\},$
pick a vector $d'\not =0 $  from this null space. Since $\Phi_{m,
S}$ has a full-column rank, we must have that   $ \left[
                          \begin{array}{c}
                             \Phi_{J_+\backslash T', S} \\
                            \Phi_{ J_-\backslash T'', S}
                          \end{array}
                        \right]d' \not =0.$
 So we can  continue to update the
components of $x'$ by setting $x'_S \leftarrow x'_S+\lambda' d'$ and
keeping $x'_i =0$ for $i\notin S,$ where $\lambda'$ is chosen such
that $x'_S+\lambda' d'$ is still feasible to (\ref{1bit-basis}) and
one of the inactive constraints at the current point $x'$  becomes
active at $x'_S+\lambda' d'.$  Thus the index sets $T'$ and
$T''$ for active constraints  are further augmented.

Since $\Phi_{m, S}$ has a full-column rank, after repeating the
above process a finite number of times,
     we  stop at a point, denoted still by $x',$ at which
                            ${\cal N} (\left[
                          \begin{array}{c}
                             \Phi_{T', S} \\
                            \Phi_{ T'', S} \\
                            \Phi_{J_0, S}  \\
                          \end{array}
                        \right])  =\{0\},$ i.e., $\left[
                          \begin{array}{c}
                             \Phi_{T', S} \\
                            \Phi_{ T'', S} \\
                            \Phi_{J_0, S}  \\
                          \end{array}
                        \right] $  has a full-column rank.
 Note that $\textrm{supp} (x') \subseteq \textrm{supp}(x^*) $ is always maintained in the above process.
Define the  sets
\begin{equation}\label{TT00}
   T_1
 = \widetilde{ {\cal A}}_+ (x'),
 ~    T_2 =
  \widetilde{
{\cal A}}_- (x') .
\end{equation}
Thus $T_1 \subseteq J_+$ and $T_2\subseteq J_-.$  By the
construction of $x',$ we see that ${\cal A}(x') \not
=\emptyset.$ Thus $(T_1, T_2)$ given by (\ref{TT00}) satisfies that
$T_1\cup T_2 \not = J_+ \cup J_-.$

We now further prove that $x'$ must be the unique optimal solution
to the 1-bit basis pursuit (\ref{1bit-basis}). By Theorem
\ref{Ness-Suff}, it is sufficient to prove that $\Phi^T$ has the
RRSP at $x'$ and the matrix
  $$H(x') = \left[
         \begin{array}{cc}
           \Phi_{\mathcal{A}(x')\cap J_+, S'_+} & \Phi_{\mathcal{A}(x')\cap J_+, S'_-} \\
           \Phi_{\mathcal{A}(x')\cap J_-, S'_+} & \Phi_{\mathcal{A}(x')\cap J_-, S'_-} \\
           \Phi_{J_0,S'_+} & \Phi_{J_0,S'_-}
           \end{array}
       \right]
$$ has a full-column rank, where $S'_+=\{i: x'_i >0\}$ and $S'_- =\{i: x'_i<0\}.$

  Indeed, let $S'_+, S'_-, T_1 $ and $T_2$ be
defined as above. Since $x'$ is consistent with $y$ and satisfies that $\textrm{supp} (x') \subseteq \textrm{supp}(x^*), $ we see that $(S'_+, S'_-) \in P(y) $
satisfying $S'= S'_+\cup S'_- \subseteq  S. $  Since $\left[
                          \begin{array}{c}
                             \Phi_{T', S} \\
                            \Phi_{ T'', S} \\
                            \Phi_{J_0, S}  \\
                          \end{array}
                        \right]      $  has a full-column rank,   $  \left[
                          \begin{array}{c}
                             \Phi_{T', S'} \\
                            \Phi_{ T'', S'} \\
                            \Phi_{J_0, S'}  \\
                          \end{array}
                        \right]       $  must have a full-column rank.
Note that
{\small \begin{equation} \label{Union} T'=  J_+ \setminus T_1 ={\cal
A} (x')\cap J_+ ,  ~~ T''= J_- \setminus T_2 ={\cal A} (x')  \cap
J_- .
\end{equation} }
Thus $H(x') =\left[\begin{array}{c}
     \Phi_{J_+\backslash T_1,S'} \\
     \Phi_{J_-\backslash T_2,S'} \\
     \Phi_{J_0,S'}
   \end{array}\right]$  has a full-column rank.

Since $\Phi^T$ has the S-RRSP of order $k$ with respect to $y, $
there exists a vector $\eta\in \mathcal{R}(\Phi^T)$  and $w\in
\mathcal{F}(T_1,T_2)$ satisfying that $\eta=\Phi^T w$  and $
  \eta_i=1 $ for $ i\in S'_+,$ $
  \eta_i=-1 $ for  $ i\in S'_-,$ and
  $|\eta_i|<1  $  otherwise.
The set $ \mathcal{F}(T_1,T_2) $ is defined as (\ref{F4}).  From
(\ref{TT00}), we see that the conditions $w_{T_1\cup T_2}=0 $ in
(\ref{F4}) coincides with the condition $w_i=0\textrm{ for }
i\in{\tilde{\mathcal{A}}_+(x')}
  \cup {\tilde{\mathcal{A}}_-(x')}.$
This, together with (\ref{Union}),  implies that
$\mathcal{F}(T_1,T_2)$ coincides with ${\cal F} (x')$ defined as
(\ref{FFFF}). Thus the RRSP of $\Phi^T$  at $x' $ holds (see
Definition \ref{DefRRSP}). This, together with the full-column-rank
property of
 $H(x') ,$ implies that $x'$ is the unique optimal solution to (\ref{1bit-basis}).

 Furthermore, suppose that $x^* $ is a $k$-sparse signal and $x^*$ is a sparsest signal
 consistent with $y.$  Since  $x'$ is also consistent with $y$,
  it follows from   $\textrm{supp} (x') \subseteq \textrm{supp} (x^*)$
  that $\textrm{supp} (x') = \textrm{supp} (x^*).$ So $x'$ is also a sparsest vector consistent with $y.$
   From the aforementioned construction process of $x',$
 it is not difficult to see  that the updating scheme  $x'_S \leftarrow x_S'+\lambda ' d'$
 does not change the sign of nonzero components of the vectors. In fact,
  when we vary the parameter $\lambda $ in  $ x_S'+\lambda  d'$ to determine the critical value $\lambda '$
  which
  yields new active
 constraints,
this value $\lambda '$ still ensures that the new vector
$x_S'+\lambda ' d'$ is  feasible to (\ref{1bit-basis}). If there is
a nonzero component of $x_S'+\lambda ' d'$, say the $i$th component,
holds a different sign from the corresponding nonzero component of
$x'_S,$ then by continuity and by convexity of the feasible set of
(\ref{1bit-basis}),   there is a suitable $\lambda $ lying between
zero  and $\lambda' $ such that the $i$th  component of
$x_S'+\lambda d'$ is equal to zero. Thus $x_S'+\lambda d'$ is
sparser than $x^*.$ Since  $x_S'+\lambda d'$ is also  feasible to
(\ref{1bit-basis}), it is  consistent with $y.$ This is a
contradiction as $x^*$ is a sparsest signal consistent with $y.$
Therefore,   we must have $\textrm{sign}(x') =\textrm{sign} (x^*).$
~ $\Box$


\section{Uniform sign recovery  }
\vskip 0.05in

Theorems \ref{Main36} and \ref{Main38} provide some conditions for
the nonuniform recovery of the sign of an individual $k$-sparse
signal. In this section, we develop some necessary and sufficient
conditions for the uniform recovery of the sign  of all $k$-sparse
signals through a sensing matrix $\Phi.$   Let us first define  $$
Y^k= \{y: y=\textrm{sign} (\Phi x), x\in R^n, \|x\|_0 \leq k\}.$$
For any two disjoint subsets $S_1, S_2 \subseteq \{1, \dots, n\}$
satisfying $|S_1\cup S_2|\leq k,$ there exists a $k$-sparse signal
$x$ such that $S_1 =S_+(x)$ and $ S_2=S_- (x).$ Thus any such
disjoint subsets $(S_1, S_2)$ must be in the set $P(y)$ for some
$y\in Y^k.$ We now introduce the notion of
 the N-RRSP of order $k$ which turns out to be a necessary condition
 for uniform sign recovery.

\vskip 0.05in
\begin{Def} [N-RRSP of order $k$] \label{N-RRSP} The matrix
$\Phi^T$ is said to satisfy the necessary restricted range space
property (N-RRSP) of order $k$ if  for any disjoint subsets $S_+, S_-
$ of $\{1,\dots, n\}$ with $|S|\leq k,$ where $S=S_+\cup S_-,$ there
exist  $y\in Y^k$ and  $(T_1, T_2) $ such that $(S_+, S_-)\in P(y),
$ $T_1 \subseteq J_+ (y)$, $T_2 \subseteq J_-(y),  $ $T_1 \cup T_2 \not = J_+(y) \cup  J_-(y) $ and {\small  $
\left[\begin{array}{c}
     \Phi_{J_+(y)\backslash T_1,S} \\
     \Phi_{J_-(y)\backslash T_2,S} \\
     \Phi_{J_0,S}
   \end{array}\right]  $ }  has a full-column rank,
 and there is a vector $\eta \in
{\cal R}(\Phi^T)$ satisfying the following properties:
\begin{enumerate}  \item [(i)] $
  \eta_i=1  \text{ for }i\in S_+,
 ~  \eta_i=-1 \text{ for }i\in S_-,
   ~ |\eta_i|<1 \text{ otherwise};
$
\item[(ii)]   $\eta=\Phi^T w$  for some $w\in \mathcal{F}(T_1,T_2)$ defined
by (\ref{F4})
\end{enumerate}
\end{Def}

\vskip 0.05in

The N-RRSP of order $k$ is a necessary condition for the uniform
recovery of the sign  of all $k$-sparse signals via 1-bit
measurements and basis pursuit.

\vskip 0.05in

 \begin{Thm}\label{Uniform-1} Let $\Phi \in R^{m\times n} $ be a given  matrix and assume that for any $k$-sparse signal $x^*$, the sign measurements
 $\textrm{sign} (\Phi x^*)$ can be acquired.  If the sign of  any $k$-sparse
signal $x^*$ can be exactly recovered by the 1-bit basis pursuit
(\ref{1bit-basis}) with $ J_+= \{i: \textrm{sign}(\Phi x^*)_i=1\}, $
$ J_-=\{i:\textrm{sign}(\Phi x^*)_i=-1\} $ and $ J_0 =\{i:
\textrm{sign}(\Phi x^*)_i= 0\} $   in the sense that
(\ref{1bit-basis}) admits a unique optimal solution $\widetilde{x}$
satisfying $\textrm{sign} (\widetilde{x})=\textrm{sign} (x^*), $
then  $\Phi^T $ must admit the N-RRSP of order $k.$
\end{Thm}

\emph{Proof.} Let $x^*$ be an arbitrary  $k$-sparse signal with
$S_+= \{i: x^*_i>0\} ,$ $ S_-=\{i: x^*_i <0\}$ and $S=S_+\cup S_-.$
Clearly, $|S| \leq k.$  Let $y= \textrm{sign}(\Phi x^*)$ be the
acquired measurements. Assume that $\widetilde{x}$ is the unique
optimal solution to (\ref{1bit-basis}) and $\textrm{sign}
(\widetilde{x})= \textrm{sign} (x^*)   .$  Then we see that $y\in
Y^k$, $ (S_+, S_-) \in P(y),$ and \begin{equation}
\label{set-11}(S_+(\widetilde{x}), S_-(\widetilde{x})) = (S_+,
S_-).\end{equation}  It follows from Theorem
 \ref{Ness-Suff} that the uniqueness of $\widetilde{x}$ implies that the matrix
$H(\widetilde{x}) $ admits a full-column rank and there exists a
vector $\eta \in {\cal R}(\Phi^T)$ such that
\begin{itemize}
\item[(a)] $\eta_i=1$ for $ i \in S_+(\widetilde{x})$, $\eta_i=-1$ for $ i\in S_-(\widetilde{x}),$ and
$|\eta_i|<1$ otherwise;

\item[(b)]  $\eta=\Phi^T w$ for some $w\in \mathcal{F} (\widetilde{x}) $  given as
 \begin{eqnarray*}
 & \mathcal{F}(\widetilde{x}) & =  \{w \in R^m:  w_i >  0\textrm{ for } i\in {\cal A}(\widetilde{x}) \cap
 J_+(y) , \nonumber \\  &   & ~~~~ w_i < 0 \textrm{ for } i\in {\cal A}(\widetilde{x}) \cap J_- (y), \\
  &   & ~~~~  w_i=0 \textrm{ for } i\in \tilde{\mathcal{A}}_+(\widetilde{x}) \cup \tilde{\mathcal{A}}_-(\widetilde{x})\}.
 \end{eqnarray*}
 \end{itemize}
Let
 $T_1=  \widetilde{{\cal A}}_+(\widetilde{x})
\subseteq J_+ (y) $ and $ T_2 =   \widetilde{{\cal
A}}_-(\widetilde{x}) \subseteq J_- (y). $ Since $\widetilde{x}$ is
an optimal solution to (\ref{1bit-basis}), we must have that $ {\cal
A} (\widetilde{x}) \not = \emptyset, $ which implies that $T_1 \cup
T_2 \not = J_+(y) \cup J_-(y).$ Clearly, \begin{equation}
\label{set-22} {\cal A}(\widetilde{x}) \cap
 J_+ (y) = J_+ (y) \backslash T_1,  ~  \mathcal{A}(\widetilde{x}) \cap J_-(y)  = J_- (y) \backslash
 T_2. \end{equation}
 Therefore, the full-column-rank property of $H(\widetilde{x})$ implies that {\small  $ \left[\begin{array}{c}
     \Phi_{J_+(y)\backslash T_1,S} \\
     \Phi_{J_-(y)\backslash T_2,S} \\
     \Phi_{J_0,S}
   \end{array}\right]  $ } has a full-column rank.  By (\ref{set-11}) and (\ref{set-22}),  the above properties (a) and (b)
 coincide with the properties (i) and (ii) described in Definition \ref{N-RRSP}.   By
considering all possible $k$-sparse signals $x^*, $ which yield
 all possible disjoint subsets $S_+, S_- $ of $\{1,\dots, n\} $ satisfying $|S_+\cup S_-|\leq k.$   Thus  $\Phi^T $ admits
 the N-RRSP
of order $k.$  ~$\Box $

\vskip 0.05in

 It should be pointed out that for random matrices $\Phi,$
 with probability 1 the optimal solution to the linear program (\ref{1bit-basis}) is unique.
 In fact, the non-uniqueness of optimal solutions happens only if the optimal face of the feasible
 set (which is a polyhedron)
 is parallel to the objective hyperplane, and the probability for this event is zero.
 This means that the uniqueness assumption for the optimal solution of (\ref{1bit-basis}) is
 very mild and it holds almost for sure. Thus when the sensing matrix $\Phi$ is randomly generated according to a
 probability distribution, with  probability 1 the RRSP of $\Phi^T$ at its optimal solution
 $\widetilde{x}$ holds and the associated matrix $H (\widetilde{x})$ has a full-column rank.
 The N-RRSP of order $k$   is defined based on such a mild assumption.
 Theorem 4.2 has indicated that the N-RRSP of order $k$ is  a necessary requirement for the
 uniform recovery of the sign of all $k$-sparse signals
from 1-bit measurements with  the linear program
(\ref{1bit-basis}). Using linear programs
 as decoding methods will necessarily  and inevitably yield a certain range space property
  like the RRSP (since this property results directly from the fundamental optimality condition of linear programs). From
  the study in this paper,  we conclude that if the sign   of  $k$-sparse
  signals can be exactly recovered from   1-bit measurements with a linear programming decoding method,
  then  $\Phi^T$
must satisfy the N-RRSP of order $k$ or its variants. At the moment,
it is not clear whether this necessary condition is also sufficient
for the exact sign recovery in 1-bit CS setting.

In classic CS, a sensing matrix is required to admit a  general
positioning property in order to achieve the uniform recovery of
$k$-sparse signals. This property is reflected in all concepts such
as RIP, NSP and RSP. Similarly, in order to the achieve the uniform
 recover of the sign of $k$-sparse signals in 1-bit CS setting, the  matrix
should admit a certain general positioning property as well. Since
N-RRSP is a necessary property for uniform sign recovery, a
sufficient sign recovery condition can be developed  by slightly
enhancing this necessary property,  i.e., by considering all
possible sign measurements $y\in Y^k$ together with the pairs $(T_1
, T_2)$ described in Definition \ref{N-RRSP}. This naturally leads
to the next definition.

\vskip 0.05in
\begin{Def} [S-RRSP of order $k$] \label{S-RRSP} The matrix
$\Phi^T$ is said to satisfy the sufficient restricted range space
property (S-RRSP) of order $k$ if for any disjoint subsets   $(S_+,
S_-) $  of $\{1, \dots, n\} $ with $|S|\leq k,$ where $S=S_+\cup
S_-,$ and for any  $y\in Y^k$ such that $(S_+, S_-)\in P(y) $, there
exist $T_1$ and $T_2$ such that  $T_1 \subseteq J_+ (y),$   $T_2 \subseteq J_-(y),  $ $ T_1 \cup T_2 \not  = J_+(y) \cup J_-(y) $ and
{\small $ \left[\begin{array}{c}
     \Phi_{J_+(y)\backslash T_1,S} \\
     \Phi_{J_- (y) \backslash T_2,S} \\
     \Phi_{J_0,S}
   \end{array}\right] $ }  has a full-column rank,
  and for any such a pair  $(T_1, T_2), $ there is a vector $\eta \in
{\cal R}(\Phi^T)$ satisfying the following properties:
\begin{enumerate} \item [(i)] $
  \eta_i=1  \text{ for }i\in S_+,
 ~  \eta_i=-1 \text{ for }i\in S_-,
   ~ |\eta_i|<1 \text{ otherwise};
$
\item[(ii)]   $\eta=\Phi^T w$  for some $w\in \mathcal{F}(T_1,T_2) $ defined
by (\ref{F4}).
\end{enumerate}
\end{Def}


\vskip 0.05in

The above concept taking into account all possible vectors $y$ is
stronger than Definition \ref{SRRSP-y}. If a matrix has the S-RRSP of
order $k,$ it must have the S-RRSP of order $k$ with respect to any
individual vector $y \in Y^k.$
The S-RRSP of order $k$  makes it possible to  recover the sign of
all $k$-sparse signals from 1-bit measurements with
(\ref{1bit-basis}), as shown in the next theorem.

\vskip 0.05in

\begin{Thm}\label{Main2}  Suppose that $\Phi^T$ has the S-RRSP of order $ k  $ and that for any  $k$-sparse
signal $x^*,$ the sign measurements $   \textrm{sign}(\Phi x^*)$ can
be acquired.  Then the 1-bit basis pursuit (\ref{1bit-basis}) with $
J_+= \{i: \textrm{sign}(\Phi x^*)_i=1\}, $ $
J_-=\{i:\textrm{sign}(\Phi x^*)_i=-1\} $ and $ J_0 =\{i:
\textrm{sign}(\Phi x^*)_i= 0\} $  has a unique optimal solution
$\widetilde{x} $ satisfying that $\textrm{supp} (\widetilde{x})
\subseteq \textrm{supp}(x^*). $ Furthermore,  for any $k$-sparse
signal $x^*$ which is a sparsest signal satisfying
\begin{equation}   \label{m-system}\textrm{sign}(\Phi x)= \textrm{sign} (\Phi x^*),\end{equation}   the sign of
$x^*$ can be exactly recovered by (\ref{1bit-basis}), i.e.,   the
unique optimal solution $ \widetilde{x} $ of (\ref{1bit-basis})
satisfies that $\textrm{sign} (\widetilde{x})=\textrm{sign} (x^*).$

\end{Thm}

\vskip 0.05in

\emph{Proof.} Let $x^*$  be an arbitrary $k$-sparse signal, and let
 measurements  $y=\textrm{sign} (\Phi x^*)$ be taken,
  which determines a partition $(J_+, J_-,
J_0)$ of $\{1, \dots, m\}$ as (\ref{JJJJ}). Since $\Phi^T$ has the
S-RRSP of order $k$, this implies that $\Phi^T$ has the S-RRSP of
order $k$ with respect to this  vector $y .$ By Theorem
\ref{Main38}, the problem (\ref{1bit-basis}) has a unique optimal
solution, denoted by $\widetilde{x},$ which satisfies that
  $\textrm{supp}(\widetilde{x}) \subseteq \textrm{supp} (x^*).$
Furthermore,  if $x^*$ is a sparsest signal satisfying the system
(\ref{m-system}), then by Theorem \ref{Main38} again, we must have
that $\textrm{sign}(\widetilde{x}) =  \textrm{sign} (x^*),$ and
hence  the sign of $x^*$ can be  exactly recovered by
(\ref{1bit-basis}).
 ~  $\Box$

  \vskip 0.05in

The above theorem indicates that under the S-RRSP of order $k$ if
$x^*$ is a sparsest solution to (\ref{m-system}),   then the sign of
$x^*$  can be exactly recovered by  (\ref{1bit-basis}). If $x^*$ is
not a sparsest solution to (\ref{m-system}), then at least part of
the support of $x^*$ can be exactly recovered by (\ref{1bit-basis})
in the sense that $\textrm{supp} (\widetilde{x}) \subseteq
\textrm{supp} (x^*),$ where $\widetilde{x}$ is the optimal solution
to (\ref{1bit-basis}).

   The study in this paper indicates that  the models (\ref{l0LP})
    and (\ref{1bit-basis})   make  it possible to establish a sign recovery theory for
     $k$-sparse signals from 1-bit measurements.
  It is worth noting that these models  can  also make it possible to extend
     reweighted $\ell_1$-algorithms
(e.g., \cite{CWB08, ZL2012, SS2013, ZK2014}) to  1-bit CS problems.

 The RIP and NSP recovery conditions are  widely assumed in classic CS
 scenarios.
 Recent study has shown that it is NP-hard to compute the RIP and NSP constants of a
 given matrix (\cite{TP14, BDMS13}).   The RSP recovery condition introduced in \cite{YBZ2013}
 is equivalent to the NSP since both are the necessary and sufficient condition for the uniform recovery of all $k$-sparse
 signals. The NSP characterizes the uniform recovery from the perspective of
the null space of a sensing matrix,
 while the RSP characterizes the uniform recovery from its orthogonal space, i.e., the range space of
 a transposed sensing matrix.
 So it is also difficult to certify the RSP of a given matrix.
 Clearly, the
   N-RRSP and S-RRSP are more complex than the standard RSP, and thus they  are hard to certify as well.
    Note that the existence of a matrix with
   the RSP follows directly from the fact that any matrix with
 RIP of order $2k$ or NSP of order $2k$ must admit the RSP of order
 $k$ (see \cite{YBZ2013}).  In 1-bit CS setting, however, the analogous theory are still
 underdevelopment.
 The existence analysis of a S-RRSP matrix has not yet properly addressed at the current
 stage.

\section{Proof of Theorem 3.2}
We now prove Theorem \ref{Ness-Suff} which provides a complete
characterization  for the uniqueness of solutions to the 1-bit basis
pursuit (\ref{1bit-basis}).
 We start by developing necessary conditions.

\subsection{Necessary condition (I):  Range space property}

By introducing   $u, v, t\in R^n_+,$ where $t$ satisfies
that $|x_i|\leq t_i$ for $i=1, \dots,n,$   then
(\ref{8888}) can be written as the linear program
  \begin{eqnarray}\label{9999}
   & \min & e^T t  \nonumber  \\
         &  \textrm{s.t.}& x +u= t,
         ~ -x+v= t,
         ~\Phi_{J_+,n} x -\alpha=  e_{J_+}, \nonumber  \\
         &  &  \Phi_{J_-,n} x +\beta=-  e_{J_-},
        ~\Phi_{J_0,n} x=0,  \\
        &      & (t,~u,~v,~\alpha,~\beta)\geq 0. \nonumber
\end{eqnarray}
Clearly,  we have the following statement.

 \vskip 0.05in

\begin{Lem}\label{LemLP2LP3}  (i)  For any optimal solution $(x^*,$ $t^*,u^*,$ $v^*,\alpha^*,$ $\beta^*)$
of   (\ref{9999}), we have that  $ t^*=|x^*|, $ $ u^*= |x^*|-x^* , $
  $ v^* =|x^*|+x^*  $ and $(\alpha^*, \beta^*)$ is given by
(\ref{Lemma31}).  (ii)   $x^*$   is the unique optimal solution to
  (\ref{1bit-basis}) if and only if
$(x,t,u,v,\alpha,\beta)=(x^*,|x^*|,|x^*|-x^*,|x^*|+x^*,\alpha^*,\beta^*)$
is the unique optimal solution to (\ref{9999}), where $(\alpha^*,
\beta^*)$  is given by
 (\ref{Lemma31}).
\end{Lem}

\vskip 0.05in

Any linear program can be written in the form $\min\{c^Tz: Az=b,
z\geq 0\},$ to which the Lagrangian dual problem is given by
$\max\{b^T y: A^T y\leq c\}$ (see, e.g., \cite{D63}).
 So it is very easy to verify that the   dual problem of
(\ref{9999}) is given as
{\small \begin{align}
&\textrm{(DLP)}& \max~ &~    e_{J_+}^T h_3-  e_{J_-}^T h_4 \nonumber &   \\
            &  &   \textrm{s.t.}~~      &  h_1-h_2+(\Phi_{J_+,n})^Th_3+(\Phi_{J_-,n})^Th_4 \nonumber \\
 & &  & +(\Phi_{J_0,n})^Th_5 = 0,& \nonumber \\
            &  &              &-h_1-h_2\leq e,& \label{cc1} \\
            &  &              &h_1\leq0,&  \label{cc2} \\
            &  &              &h_2\leq0,&  \label{cc3}\\
            &  &              &-h_3\leq0,&  \label{cc4} \\
            &  &              & h_4\leq0.&  \label{cc5}
\end{align} }
The (DLP) is always feasible in the sense that there exists a point,
for instance, $(h_1, \dots, h_5)= (0, \dots, 0),$ satisfies all
constraints. Furthermore, let $ s^{(1)},\dots, s^{(5)}$ be the
nonnegative slack variables associated with the constraints
(\ref{cc1}) through (\ref{cc5}), respectively. Then   (DLP) can be
also written as
{\small \begin{eqnarray}
 & \max~ &~    e_{J_+}^T h_3-  e_{J_-}^T h_4 \nonumber   \\
               &   \textrm{s.t.}~~      &  h_1-h_2+(\Phi_{J_+,n})^Th_3+(\Phi_{J_-,n})^Th_4 \nonumber \\
 &    & +(\Phi_{J_0,n})^Th_5 = 0,   \label {cc0} \\
              &              &  s^{(1)}-h_1-h_2= e,\label{slack1}\\
               &              & s^{(2)}+h_1=0,\label{slack2}\\
              &              &  s^{(3)}+h_2=0,\label{slack3}\\
               &              & s^{(4)}-h_3=0,\label{slack4}\\
              &              & s^{(5)}+h_4=0,\label{slack5}\\
 &    & s^{(1)}, \dots ,  s^{(5)} \geq 0. \nonumber
\end{eqnarray}}
We now prove that if $x^*$ is the unique optimal  solution to
 (\ref{1bit-basis}), the range space   $\mathcal{R}(\Phi^T)$  must satisfy some properties.

\vskip 0.05in

\begin{Lem} \label{RSPThm}
 If $x^*$ is the unique optimal solution to  (\ref{1bit-basis}), then there exist  vectors $h_1,h_2 \in
R^n$ and
  $ w\in R^m$ satisfying
{\small \begin{eqnarray} \label{RRSP-CON}
\left\{ {\small
\begin{array}{ll}
h_2-h_1=\Phi^Tw, &   \\
(h_1)_i=-1,~(h_2)_i=0 &  \text{for } x_i^*>0, \\
(h_1)_i=0,~(h_2)_i=-1 &  \text{for } x_i^*<0,\\
 (h_1)_i,(h_2)_i<0,(h_1+h_2)_i>-1   & \small \text{for } x^*_i=0,\\
\begin{array} {cl}
 w_i> 0  &  \textrm{ for }i\in \mathcal{A}(x^*)\cap J_+,  \\
w_i<  0  & \textrm{  for }i\in \mathcal{A}(x^*)\cap J_-,  \\
w_i=0  & \textrm{ for } i\in \tilde{\mathcal{A}}_+(x^*) \cup
\tilde{\mathcal{A}}_-(x^*).
\end{array} & \\
\end{array}
} \right.
\end{eqnarray} }

\end{Lem}

\emph{Proof. }   Assume that $x^*$ is the unique optimal solution to
  (\ref{1bit-basis}).   By Lemma \ref{LemLP2LP3},
\begin{eqnarray} \label{sol**}
(x,t,u,v,\alpha,\beta)=(x^*,|x^*|,|x^*|-x^*,|x^*|+x^*,\alpha^*,\beta^*)
\end{eqnarray}
is the  unique optimal solution to    (\ref{9999}),  where
$(\alpha^*, \beta^*) $ is given by (\ref{Lemma31}). By the strict
complementarity theory of linear programs (see, e.g., Goldman and
Tucker \cite{GT56}) ,
 there exists a solution $(h_1, \dots, h_5)$ of (DLP)
   such that the associated
     vectors $ s^{(1)},\dots, s^{(5)} $ determined by
   (\ref{slack1})--(\ref{slack5}) and the vectors $(t, u,v, \alpha, \beta) $ given by (\ref{sol**})   are
 strictly complementary, i.e.,
these vectors satisfy the  conditions
\begin{equation} \label{strict1}
 t^Ts^{(1)}=u^Ts^{(2)}= v^Ts^{(3)}=\alpha^Ts^{(4)}= \beta^Ts^{(5)}=0
\end{equation}
and
\begin{equation}\label{strict2}
 \left\{\begin{array}{l} t+s^{(1)}>0, ~ u+s^{(2)}>0, ~ v+s^{(3)}>0, \\
 \alpha+s^{(4)}>0, ~\beta+s^{(5)}>0.
 \end{array} \right.
\end{equation}
For the above-mentioned solution $(h_1, \dots, h_5)$ of (DLP), let
$w\in R^m $ be the vector defined by $w_{J_+} = h_3, w_{J_-}=h_4,$
and $ w_{J_0}=h_5.$ Then  it follows from (\ref{cc0}) that
\begin{equation}\label{RangeCond}
h_2-h_1= (\Phi_{J_+,n})^Th_3+(\Phi_{J_-,n})^Th_4+(\Phi_{J_0,n})^Th_5
=\Phi^T w.
\end{equation}
From (\ref{sol**}),   we see that the solution  of  (\ref{9999})
satisfies the following properties:
\begin{eqnarray*}
 \begin{array}{cccl}
& t_i=x_i^*>0, ~ u_i=0, ~ v_i=2x^*_i>0 &\text{for   } x_i^*>0,\\
& t_i=|x_i^*|>0, ~ u_i=2|x_i^*|>0,  ~ v_i=0 &\text{for  } x_i^*<0,\\
& t_i=0, ~  u_i=0, ~ v_i=0 &\text{for }  x^*_i=0.
\end{array}
\end{eqnarray*}
Thus, from (\ref{strict1}) and (\ref{strict2}), it follows that
\begin{eqnarray*}
\begin{array}{llll}
s^{(1)}_i=0,& s^{(2)}_i>0,& s^{(3)}_i=0 &\text{for   } x_i^*>0,\\
s^{(1)}_i=0,& s^{(2)}_i=0,& s^{(3)}_i>0  &\text{for   } x_i^*<0,\\
s^{(1)}_i>0,& s^{(2)}_i>0,& s^{(3)}_i>0  &\text{for }  x^*_i=0.
\end{array}
\end{eqnarray*}
From  (\ref{slack1}), (\ref{slack2}) and (\ref{slack3}), the above
relations imply that
\begin{eqnarray*}
\begin{array}{llll}
(h_1+h_2)_i=-1,& (h_1)_i<0,&(h_2)_i=0  &\text{ for   } x_i^*>0,\\
(h_1+h_2)_i=-1,& (h_1)_i=0,&(h_2)_i<0  &\text{ for   } x_i^*<0,\\
(h_1+h_2)_i>-1,& (h_1)_i<0,&(h_2)_i<0  &\text{ for }  x^*_i=0.
\end{array}
\end{eqnarray*}
From  (\ref{slack4}) and (\ref{slack5}), we see that $
s^{(4)}=h_3\geq 0 $ and $ s^{(5)}=-h_4\geq 0. $ Let $\pi(\cdot)$ and
$\varrho (\cdot) $ be defined as (\ref{pi}) and (\ref{varrho}),
respectively. It follows from (\ref{Lemma31}), (\ref{strict1}) and
(\ref{strict2}) that
   \begin{eqnarray*} (h_3)_{\pi(i)} & = &  s^{(4)}_{\pi(i)} >0
\textrm{ for } i\in \mathcal{A}(x^*)\cap J_+, \\
  (h_3)_{\pi(i)} & = &
s^{(4)}_{\pi(i)}=0 \textrm{ for }  i\in \tilde{\mathcal{A}}_+(x^*),
\\
  (-h_4)_{\varrho(i)}  & = &  s^{(5)}_{\varrho(i)} >0 \textrm{ for   } i\in \mathcal{A}(x^*)\cap
 J_-, \\ (-h_4)_{\varrho(i)} & =  & s^{(5)}_{\varrho(i)} =0 \textrm{  for  }  i\in
\tilde{\mathcal{A}}_-(x^*).
\end{eqnarray*}
By the definition of $w$ (i.e., $w_{J_+} = h_3 ,$ $ w_{J_-}=h_4 $
and $w_{J_0}=h_5$), the above conditions imply that
{\small \begin{eqnarray*} w_i & = &  (h_3)_{\pi(i)} >0
\textrm{ for } i\in \mathcal{A}(x^*)\cap J_+, \\
 w_i  & = &
(h_3)_{\pi(i)}=0 \textrm{ for } i\in \tilde{\mathcal{A}}_+(x^*),
\\
  w_i  & = &  (h_4)_{\varrho(i)} <0 \textrm{ for  } i\in \mathcal{A}(x^*)\cap
 J_-, \\
 w_i & = & (h_4)_{\varrho(i)} =0 \textrm{  for  }  i\in
\tilde{\mathcal{A}}_-(x^*).
\end{eqnarray*} }
 Thus,
$h_1,h_2 $ and $ w $ satisfy (\ref{RangeCond}) and the  properties:
{\small \begin{align*}
\begin{array}{cl}
(h_1)_i=-1,~(h_2)_i=0 &  \text{ for } x_i^*>0, \\
(h_1)_i=0,~(h_2)_i=-1 &  \text{ for }~x_i^*<0,\\
(h_1)_i,(h_2)_i<0, ~(h_1+h_2)_i>-1 &\text{ for }~x^*_i=0,\\
w_i>0&\textrm{ for }i\in \mathcal{A}(x^*)\cap J_+,\\
w_i=0&\textrm{ for } i\in \tilde{\mathcal{A}}_+(x^*),\\
w_i<0&\textrm{ for }i\in \mathcal{A}(x^*)\cap J_-,\\
w_i=0&\textrm{ for }i\in \tilde{\mathcal{A}}_-(x^*).
\end{array}
\end{align*} }
Therefore,   condition (\ref{RRSP-CON}) is a necessary condition for
$x^*$ to be the unique optimal solution to (\ref{1bit-basis}). ~ $
\Box $

\vskip 0.05in

It should be pointed out that the uniqueness  of $x^*$  implies that
$x^*$  is the  strictly complementary solution. This leads to the
condition (\ref{RRSP-CON}) in  which all inequalities hold strictly.
If $x^*$ is not
  the unique optimal solution of (\ref{1bit-basis}), then  $x^*$ is not necessarily a strictly complementary
  solution, and thus (\ref{RRSP-CON}) does not necessarily hold.
  We now present an equivalent
statement for (\ref{RRSP-CON}) as follows.

\vskip 0.05in

\begin{Lem}\label{Lemyeta}
Let $x^*\in R^n$ be  a given  vector satisfying the constraints of
(\ref{1bit-basis}). There exist vectors $h_1,h_2$ and $w$ satisfying
(\ref{RRSP-CON}) if and only if there exists a vector
$\eta\in\mathcal{R}(\Phi^T) $ satisfying the following two
conditions:
\begin{itemize}
\item[(i)] $\eta_i=1$ for $x^*_i>0$, $\eta_i=-1$ for $x^*_i<0,$ and
$|\eta_i|<1$ for $x^*_i=0$;
\item[(ii)]  $\eta=\Phi^T w$ for some $w\in \mathcal{F} (x^*) $   defined as (\ref{FFFF}).
\end{itemize}
\end{Lem}

It is straightforward to verify this lemma. Its proof is omitted here.
 By Definition \ref{DefRRSP}, Combining Lemmas
\ref{RSPThm} and   \ref{Lemyeta} yields the following result.

\vskip 0.05in

\begin{Cor} \label{Cor54}
If $x^*$ is the unique optimal solution to (\ref{1bit-basis}), then
the RRSP of $\Phi^T$ at $x^*$ holds.
\end{Cor}

\vskip 0.05in

The RRSP at $x^*$  is not sufficient to ensure the uniqueness of
$x^*.$   We need to develop another necessary condition (called the
full-column-rank property).

\subsection{Necessary condition (II):  Full column rank}

 Assume
that $x^*$ is the unique optimal solution to (\ref{1bit-basis}).
Denote still by $ S_+=\{i: x^*_i>0\}$ and $ S_-=\{i: x^*_i<0\}. $ We
have the following lemma.



\vskip 0.05in

\begin{Lem}\label{Lemma37} If $x^*$ is the unique optimal solution
to (\ref{1bit-basis}), then  $H(x^*),$ defined by (\ref{FRPmatrix}),
has a full-column rank.
\end{Lem}

\vskip 0.05in

{\it Proof.} Assume the contrary that $H(x^*)$ has linearly
dependent columns.
Then there exists a   vector $d= \left[
                                   \begin{array}{c}
                                     u \\
                                     v \\
                                   \end{array}
                                 \right]
\neq 0, $ where $ u\in R^{|S_+|}$ and $ v\in R^{|S_-|},$   such that
$ H(x^*) d= 0.
$
Since $x^*$ is the unique optimal solution to (\ref{1bit-basis}),
there exist nonnegative $\alpha^*$ and $\beta^*,$ determined  by
(\ref{Lemma31}), such that $(x^*,\alpha^*,\beta^*)$ is the unique
optimal solution to (\ref{8888}) with the least objective value
$\|x^*\|_1$. Note that $(x^*, \alpha^*, \beta^*)$ satisfies
$$ \Phi_{J_+, n} x^* -\alpha^* =
e_{J_+}, ~ \Phi_{J_-, n} x^* +\beta^* = -  e_{J_-},  ~
\Phi_{J_0, n} x^* = 0. $$
Similar to the proof of Lemma \ref{Lemma41}, eliminating the zero
components of $x^*$, $\alpha^*$ and $\beta^*$ from the above system
 yield the same system as (\ref{system1}).
Similarly, we define $x(\lambda) \in R^n$ as $ x(\lambda)_{S_+}=
x^*_{S_+}+\lambda u, $ and $  x(\lambda)_{S_-}= x^*_{S_-} + \lambda
v, $ and $ x(\lambda)_i=0 \textrm{ for }i \notin S_+ \cup S_-.
$
We see that for all sufficiently small $|\lambda|,$
$(x(\lambda)_{S_+},  x(\lambda)_{S_-})$   satisfies the conditions
(\ref{H1})--(\ref{H3}). In other words, there exists a small number
$\delta >0$ such that for any $\lambda\neq 0$ with  $ |\lambda|\in
(0, \delta), $ the vector $  x(\lambda) $ is feasible to
(\ref{1bit-basis}). In particular, choose $\lambda^* \not=0 $ such
that $| \lambda^* |\in (0,\delta), x^*_{S_+}+\lambda^* u >0,
x^*_{S_-} + \lambda^* v>0 $ and
\begin{equation} \label {ll}
\lambda^*(e_{S_+}^Tu-e_{S_-}^Tv)\leq 0.
\end{equation}
Then we see that $ x(\lambda^*)
 \not= x^* $ since $\lambda^* \not=0 $ and $(u,v) \not =0. $
Moreover, we have
\begin{eqnarray*}
\| x(\lambda^* )\|_1&=&e_{S_+}^T (x^*_{S_+}+\lambda^* u)-e_{S_-}^T (x^*_{S_-}+\lambda^* v),\\
               &=&e_{S_+}^T x^*_{S_+}-e_{S_-}^Tx^*_{S_-}+\lambda^* e_{S_+}^T u -\lambda^* e_{S_-}^Tv,\\
               &=&\|x^*\|_1+\lambda^* (e_{S_+}^Tu -e_{S_-}^Tv) \\ & \leq  & \|x^*\|_1,
\end{eqnarray*}
where the inequality follows from (\ref{ll}). As $\|x^*\|_1$ is the
least objective value of  (\ref{1bit-basis}), it implies that
$x(\lambda^*)$ is also an optimal solution to this problem,
contradicting to the uniqueness of $x^*. $ Hence, $H(x^*)$ must have
a full-column rank. ~ $ \Box $

   Combining Corollary \ref{Cor54} and Lemma \ref{Lemma37}  yields the desired necessary conditions.

\vskip 0.05in

\begin{Thm}\label{necc}
If $x^*$ is the unique optimal solution to  (\ref{1bit-basis}), then
$ H (x^*),  $ given  by (\ref{FRPmatrix}),
 has a full-column rank and
  the RRSP of $\Phi^T$ at $x^*$ holds.
\end{Thm}

\subsection{Sufficient conditions}

We now prove that the converse  of Theorem \ref{necc}
is also valid, i.e.,   the RRSP of $\Phi^T$ at $x^*$ combined with
the full-column-rank property of $H(x^*)$  is  a sufficient
condition for the uniqueness of   $x^*. $     We start with a
property of (DLP).

\vskip 0.05in

\begin{Lem}\label{LemDualOpt}
Suppose that $x^*$ satisfies the constraints of (\ref{1bit-basis}).
If the vector $(h_1,h_2,w)\in R^n \times R^n \times R^m $ satisfies
that
{\small \begin{equation} \label {Lemma310}
\left\{
\begin{array}{ll}
(h_1)_i=-1,~(h_2)_i=0 & \text{ for }x^*_i>0,\\
(h_1)_i=0,~(h_2)_i=-1 & \text { for }x^*_i<0, \\
(h_1)_i<0, ~(h_2)_i<0, ~(h_1+h_2)_i>-1  & \text{ for }x^*_i=0, \\
h_2-h_1 = \Phi^T w,\\
w_{J_+} \geq 0,\\
w_{J_-}\leq 0,  \\
w_i=0\textrm{ for }~i\in\tilde{\mathcal{A}}_+(x^*)
 \cup \tilde{\mathcal{A}}_-(x^*),
\end{array}\right.
\end{equation} }
then the vector $(h_1,h_2,h_3,h_4,h_5),$ with $h_3=w_{J_+},
h_4=w_{J_-}$ and $ h_5=w_{J_0},$  is an optimal solution to (DLP)
and $x^*$ is an optimal solution to (\ref{1bit-basis}).
\end{Lem}

\vskip 0.05in

 This lemma follows directly from the optimality theory of linear programs by verifying  that the dual optimal value at $(h_1,h_2,h_3,h_4,h_5)$ is equal to $\|x^*\|_1.$
The proof is omitted.
We now prove the desired sufficient condition for the uniqueness of optimal
solutions of (\ref{1bit-basis}).

\vskip 0.05in

\begin{Thm}\label{suff}
Let $x^*$ satisfy the constraints of the problem (\ref{1bit-basis}).
If the RRSP of $\Phi^T$ at $x^*$  holds and $H(x^*),$ defined by
(\ref{FRPmatrix}), has a full-column rank, then $x^*$ is the unique
optimal solution to   (\ref{1bit-basis}).
\end{Thm}

\vskip 0.05in

{\it Proof.} By the assumption of the theorem, the RRSP of $\Phi^T$
at $x^*$  holds.  Then by Lemma \ref{Lemyeta}, there exists a vector
$(h_1,h_2,w)\in R^n\times R^n\times R^m$ satisfying
(\ref{RRSP-CON}), which implies that   condition (\ref{Lemma310})
holds.  As $x^*$ is feasible to (\ref{1bit-basis}), by Lemma
\ref{LemDualOpt}, $(h_1,h_2,h_3,h_4,h_5)$ with $h_3=w_{J_+},
h_4=w_{J_-}$ and $ h_5=w_{J_0}$  is an optimal solution to (DLP). At
this solution, let the slack vectors $s^{(1)},\dots,s^{(5)}$ be
given as
 (\ref{slack1})--(\ref{slack5}).  Also, from Lemma
\ref{LemDualOpt}, $x^*$ is an optimal  solution to
(\ref{1bit-basis}). Thus  by Lemma \ref{LemLP2LP3},
             $(x,t,u,v,\alpha,\beta)=
(x^*,|x^*|,|x^*|-x^*,|x^*|+x^*,\alpha^*,\beta^*), $ where
$(\alpha^*, \beta^*) $  is given by (\ref{Lemma31}), is an  optimal
solution to   (\ref{9999}).  We now further show that $x^*$ is the
unique solution to (\ref{1bit-basis}).

The vector $(x^*, \alpha^*, \beta^*)$ satisfies the system
 $   \Phi_{J_+, n} x^* -\alpha^* =
e_{J_+},   \Phi_{J_-, n} x^* +\beta^* = -  e_{J_-} $ and $
\Phi_{J_0, n} x^* = 0. $
   As shown in the proof of Lemma \ref{Lemma37},
   removing the zero components of $ (x^*, \alpha^*, \beta^*) $ from the above system yields
  \begin{equation}\label{SL1} H(x^*) \left[
                     \begin{array}{c}
                        x^*_{S_+} \\
                        x^*_{S_-}
                     \end{array}
                   \right] =   \left[
                               \begin{array}{c}
                                  e_{\mathcal{A}(x^*)\cap J_+} \\
                                 -  e_{\mathcal{A}(x^*)\cap J_-} \\
                                 0
                               \end{array}
                             \right].
                   \end{equation}
Let $(\widetilde{x},
\widetilde{t},\widetilde{u},\widetilde{v},\widetilde{\alpha},\widetilde{\beta})
$ be an arbitrary optimal solution to   (\ref{9999}).  By Lemma
\ref{LemLP2LP3}, it must hold that  $ \widetilde{t} =
|\widetilde{x}| ,\widetilde{u} =|\widetilde{x}|- \widetilde{x}  $
and
  $ \widetilde{v} =  |\widetilde{x}|+ \widetilde{x}. $ By the complementary slackness
property of linear programs (see, e.g., \cite{GT56, D63}), the nonnegative
vectors $
(\widetilde{t},\widetilde{u},\widetilde{v},\widetilde{\alpha},\widetilde{\beta})
$ and $ (s^{(1)}, \dots, s^{(5)}) $ are complementary, i.e.,
\begin{equation} \label {comp}
 \widetilde{t}^Ts^{(1)}=\widetilde{u}^Ts^{(2)}= \widetilde{v}^Ts^{(3)}
 =\widetilde{\alpha}^Ts^{(4)}= \widetilde{\beta}^Ts^{(5)}=0.
\end{equation}
As $(h_1,h_2,w)$ satisfies (\ref{RRSP-CON}), the vector $(h_1, h_2)$
satisfies that $ (h_1)_i=-1<0 $ for    $ x_i^*>0, $ $ (h_2)_i=-1<0
$ for   $ x_i^*<0 $ and that $ (h_1+h_2)_i>-1, $ $ (h_1)_i<0$ and
$ (h_2)_i<0 $  for    $ x^*_i=0. $ By the choice of $(h_1,h_2)$
and $(s^{(1)}, \dots, s^{(5)})$, we see that the following components
of slack variables are positive:
\begin{eqnarray*}
\begin{array}{cl}
s^{(1)}_i=1+(h_1+h_2)_i>0  & \text{ for }  x^*_i=0,\\
s^{(4)}_{\pi(i)}=(h_3)_{\pi(i)}= w_i>0 &\textrm{ for } i\in {\cal A}
(x^*)\cap J_+ ,\\
s^{(5)}_{\varrho(i)}=-(h_4)_{\varrho(i)}= -w_i>0 & \textrm{ for }
i\in {\cal A}(x^*)\cap J_- .
\end{array}
\end{eqnarray*}
These conditions, together with (\ref{comp}),
implies that
{\small  \begin{eqnarray}\label{eqn9}
\left\{\begin{array}{cl}
\widetilde{t}_i=0  & \text{ for }  x^*_i=0,    \\
\widetilde{\alpha}_{\pi(i)}= 0 &\textrm{ for }i\in {\cal A}
(x^*)\cap J_+ ,\\
\widetilde{\beta}_{\varrho(i)}= 0 &\textrm{ for }i\in {\cal A}
(x^*)\cap J_- .
\end{array} \right.
\end{eqnarray} }
We still use the symbol $S_+ = \{i: x^*_i >0\} $ and $ S_-= \{i:
x^*_i<0\}.$  Since  $\widetilde{ t}= |\widetilde{x}|,$  the first
relation  in (\ref{eqn9})  implies that $ \widetilde{x}_i= 0$ for
all $i \notin S_+\cup S_-. $ Note that
 $$ \Phi_{J_+, n}  \widetilde{x}  -\widetilde{\alpha}  =  e_{J_+}, ~ \Phi_{J_-, n} \widetilde{x}
+\widetilde{\beta}  = -   e_{J_-},  ~ \Phi_{J_0, n}
\widetilde{x} = 0.  $$  Since $\widetilde{x}_i =0$ for all  $ i
\notin S_+\cup S_-, $
by (\ref{XXXc}) and (\ref{eqn9}), it implies from the above system
that
  \begin{equation} \label{SL2}   H(x^*) \left[
                     \begin{array}{c}
                        \widetilde{x}_{S_+} \\
                       \widetilde{x}_{S_-}
                     \end{array}
                   \right] = \left[
                               \begin{array}{c}
                                   e_{\mathcal{A}(x^*)\cap J_+} \\
                                 -  e_{\mathcal{A}(x^*)\cap J_-} \\
                                 0
                               \end{array}
                             \right].
                 \end{equation}
By the assumption of the  theorem, the matrix $H(x^*) $
has a full-column rank.    Thus it follows from (\ref{SL1})  and
(\ref{SL2}) that $  \widetilde{x}_{S_+}=x^*_{S_+}$ and $
 \widetilde{x} _{S_-}= x^*_{S_-} $ which, together with the fact $ \widetilde{x}_i =0$ for all $i\notin S_+\cup
 S_-,$  implies that
 $ \widetilde{x} =x^*.$ By assumption,  $(\widetilde{x},
\widetilde{t},\widetilde{u},\widetilde{v},\widetilde{\alpha},\widetilde{\beta})
$ is an arbitrary optimal solution to (\ref{9999}). Thus
$(x,t,u,v,\alpha,\beta)=
(x^*,|x^*|,|x^*|-x^*,|x^*|+x^*,\alpha^*,\beta^*) $ is the unique
optimal solution to   (\ref{9999}), and hence (by Lemma
\ref{LemLP2LP3}) $x^*$ is the unique optimal solution to
 (\ref{1bit-basis}).   ~ $ \Box $

\vskip 0.05in

Combining Theorems \ref{necc} and \ref{suff} yields Theorem
\ref{Ness-Suff}.

\section{Conclusions}

Different from the classic compressive sensing, 1-bit measurements
are robust to any  small perturbation of
 a  signal. The purpose of this paper is to   show that the exact recovery of the sign of a sparse signal
  from 1-bit measurements is possible.
  We have proposed a new reformulation for the 1-bit CS
  problem. This reformulation
 makes it possible to extend the analytical tools in classic CS to 1-bit CS in order to achieve
 an analogous theory and decoding algorithms for  1-bit CS problems. Based on the fundamental Theorem
 \ref{Ness-Suff}, we
 have introduced the so-called restricted range space property (RRSP)
of a sensing matrix. This property has been used to establish
 a  connection between  sensing matrices and the sign recovery of sparse signals from 1-bit measurements.
 For nonuniform sign recovery, we have shown that
 if the transposed sensing matrix
 admits the so-called  S-RRSP of order $k$ with respect to   1-bit measurements, acquired from an individual $k$-sparse signal,
 then the sign of the signal can be exactly recovered by the proposed 1-bit basis pursuit.
 For uniform sign recovery,  we have shown that the sign of
 any  $k$-sparse signal, which is the sparsest signal  consistent with the acquired 1-bit measurements, can be exactly recovered
with   1-bit
 basis pursuit when the transposed sensing matrix
 admits the so-called S-RRSP of order $k.$


\begin{thebibliography}{999}

\bibitem{ALPV2014} A. AI, A. Lapanowski, Y. Plan and R. Vershynin,
One-bit compressed sensing with non-gaussian measurements,
 \emph{Linear Algebra Appl.}, 441, 222--239 (2014),

 \bibitem{BBR2013} S. Bahmani, P.T. Boufounos and B. Raj,  Robust
 1-bit compressive sensing via gradient support pursuit, TR2013-029,
 Mitsubishi Electric Research Lab,
 April 2013.

\bibitem{BDMS13}  A. Bandeira, E. Dobriban, D. Mixon and W. Sawin, Certifying the restricted
isometry property is hard, \emph{IEEE Trans. Inf. Theory},  59, 3448-3450 (2013).

\bibitem{B2009} P.  Boufounos, Greedy sparse signal reconstruction from sign measurements,
in \emph{Proc. 43rd Asilomar Conf. Signals Syst. Comput.}, Pacific
Grove, CA, 1305--1309 (2009).

\bibitem{B2010} P. Boufounos, Reconstruction of sparse signals from distorted randomized measurements,
in \emph{IEEE Int. Conf. Acoustics, Speech, and Signal Process.},
Dallas, TX,  3998--4001 (2010).

\bibitem{BB2008} P.  Boufounos and
R.  Baraniuk, 1-bit compressive sensing, in \emph{Proc. 42nd Annu.
Conf. Inf. Sci. Syst}, Princeton, NJ, 16--21 (2008).


\bibitem{BAU2010} A. Bourquard, F. Aguet and M. Unser,
Optical imaging using binary sensors, \emph{Opt. Express}, 18,
4876--4888 (2010).

\bibitem{BU2013} A. Bourquard and M. Unser, Binary compressed imaging,
 \emph{IEEE Trans. Image Process.}, 22, 1042--1055 (2013).

\bibitem{BED2009} A. Bruckstein, M. Elad and D. Donoho, From sparse solutions of systems of equations
    to sparse modeling of signals and images, \emph{SIAM Review}, 51, 34--81 (2009).



\bibitem{ET2005} E. Cand$\grave{e}$s and T. Tao, Decoding by linear programming,
\emph{IEEE Trans. Inf. Theory}, 51, 4203--4215 (2005).

\bibitem{ERT2006} E. Cand$\grave{e}$s, J. Romberg and T. Tao, Stable signal
recovery from incomplete and inaccurate measurements, \emph{Comm.
Pure Appl. Math.}, 59, 1207--1223 (2006).

\bibitem{ERTR2006} E. Cand$\grave{e}$s, J. Romberg and T. Tao, Robust uncertainty principles:
exact signal reconstruction from highly incomplete frequency information, \emph{IEEE Trans. Inf. Theory},
 52, 489--509 (2006).

\bibitem{CWB08} E. Cand$\grave{\textrm{e}}$s, M. Wakin and S.
Boyd, Enhancing sparsity by reweighted $\ell_1$ minimization,
\emph{J. Fourier Anal. Appl.}, 14, 877--905 (2008)

\bibitem{CDD2009} A. Cohen, W. Dahmen and R. Devore, Compressive sensing and best $k-$term approximation,
\emph{J. Amer. Math. Soc.}, 22, 211--231 (2009).



\bibitem{DPBW2014} M. Davenport, Y. Plan, E. van den Berg and M. Wootters, 1-bit matrix completion,
\emph{Information and Inference}, 3, 189--223 (2014).

\bibitem{D2006} D. Donoho, Compressed sensing, \emph{IEEE Trans. Inf. Theory}, 52, 1289--1306 (2006).

\bibitem{DE2003} D. Donoho and M. Elad, Optimally sparse representation in general (nonorthogonal)
    dictionaries via $\ell^1$ minimization, \emph{Proc. Natl. Acad. Sci. USA}, 100, 2197--2202 (2003).

 \bibitem{D63} G. Dantzig, Linear Programming and Extensions, Princeton University Press,  Princeton, NJ, 1963.

\bibitem{FR2013} S. Foucart and H. Rauhut,
 A Mathematical Introduction to Compressive Sensing,  Springer, NY, 2013.


\bibitem{F2004} J. Fuchs, On sparse representations in arbitrary
redundant bases, \emph{IEEE Trans. Inform. Theory,} 50, 1341--1344
(2004).

\bibitem{GT56} A. Goldman and A. Tucker, Theory of linear programming, in Linear Inequalities and
Related Systems (edited by H.W. Kuhn and A.W. Tucker), pp. 5397
(1956).

\bibitem{GNJN2013} S. Gopi, P. Netrapalli, P. Jain and A. Nori, One-bit compressed sensing: provable
support and vector recovery, in \emph{Proc. 30th Int. Conf. Machine
learning},  3, 154--162 (2013).

\bibitem{GNR2010} A. Gupta, R. Nowak and B. Recht, Sample complexity for 1-bit compressed sensing
and sparse classification, \emph{IEEE Int. Symp. Inf. Theory},
1553--1557 (2010).


\bibitem{JLBB2013} L. Jacques, J.  Laska, P.  Boufounos and R.  Baraniuk,
 Robust 1-bit compressive sensing via binary stable embeddings of sparse vectors,
 \emph{IEEE Trans. Inf. Theory}, 59, 2082--2102 (2013).

\bibitem{KBAU2012} U. Kamilov, A. Bourquard, A. Amini and M. Unser,
One-bit measurements with adaptive thresholds, \emph{IEEE Signal
Process. Lett.}, 19, 607--610 (2012).


\bibitem{L2011} J. Laska, Regime change: sampling rate vs
bit-depth in compressive sensing, PhD thesis, Rice university, 2011

\bibitem{LWYB2011} J. Laska, Z. Wen, W. Yin and R. Baraniuk,
    Trust, but verify: fast and accurate signal recovery from 1-bit compressive measurements,
    \emph{IEEE Trans. Signal Process.}, 59, 5289--5301 (2011).

\bibitem{LB2012} J. Laska and R. Baraniuk,
    Regime change: bit-depth versus measurement-rate in compressive sensing,
    \emph{IEEE Trans. Signal Process.}, 60, 3496--3505 (2012).

\bibitem{LRRB05} B. Le, T. Rondeau, J. Reed and C. Bostian, Analog-to-digital converters, \emph{IEEE Signal Process Mag.}, vol.22, no. 6, pp. 69-77, 2005.

\bibitem{MPD2012} A. Movahed, A. Panahi and G. Durisi,
    A robust RFPI-based 1-bit compressive sensing reconstruction algorithm,
    \emph{ IEEE Inf. Theory Workshop}, Laussane, Switzerland, 567--571 (2012).


\bibitem{PV20138} Y. Plan and R. Vershynin, One-bit compressed sensing by linear programming,
\emph{Comm. Pure Appl. Math.}, 66, 1275--1297 (2013).

\bibitem{PV20131} Y. Plan and R. Vershynin,
    Robust 1-bit compressed sensing and sparse logistic regression:
    a convex programming approach, \emph{IEEE Trans. Inf. Theory}, 59, 482--494 (2013).
    
    \bibitem{P07} M. Plumbley, On polar polytopes and the recovery of sparse representations, \emph{IEEE Trans. Infom. Theory}, 53, 3188-3195 (2007).  


\bibitem{SS2013} L. Shen and B. Suter, Blind one-bit compressive sampling, Technical Report, arXiv, 2013.

\bibitem{SG09} J. Sun and V. Goyal, Quantization for compressed sensing reconstruction, presented at the sampling Theroey Appl., Marseille, France, May 2009.

\bibitem{TP14} A. Tillmann and M. Pfetsch, The computational complexity of the restricted isometry property, the nullspace property, and related concepts in compressed sensing,
\emph{IEEE Trans. Inf. Theory},  60, 1248--1259 (2014).

\bibitem{T2004} J. Tropp, Greed is good: algorithmic results for sparse approximation,
\emph{IEEE Trans. Inf. Theory}, 50, 2231--2242 (2004).

\bibitem{W99} R. Walden, Analog-to-digital converter suvery and analysis, \emph{IEEE J. Sel. Areas Commun.}, vol. 17, no. 4, pp. 539-550, 1999.


\bibitem{YYO2012} M. Yan, Y. Yang and S. Osher, Robust 1-bit compressive sensing using adaptive outlier pursuit,
\emph{IEEE Trans. Signal Process.}, 60, 3868--3875 (2012).


\bibitem{YZ2008} Y. Zhang, Theory of compressive sensing via $\ell_1$-minimization:
a non-RIP analysis and extensions, \emph{J. Oper. Res. Soc. China},
1, 79--105 (2013).


\bibitem{YBZ2013} Y.-B. Zhao, RSP-based analysis for sparsest and least $\ell_1$-norm solutions to
underdetermined linear systems, \emph{IEEE Trans. Signal Process.},
61, 5777--5788 (2013).


\bibitem{ZL2012} Y.-B. Zhao and D. Li, Reweighted $\ell_1$-minimization for sparse solutions to
underdetermined linear systems,   \emph{SIAM J. Optim.}, 22,
1065--1088 (2012)

\bibitem{ZK2014} Y.-B Zhao and M. Ko\u{c}vara, A new computational
method for the sparsest solution to systems of linear equations,
\emph{SIAM J. Optim.},  25, 1110-1134 (2015). 


\end{thebibliography}
\end{document}